\documentclass[usenatbib]{mnras} % removed fleqn to centre the equations

\usepackage[T1]{fontenc}
\usepackage{ae,aecompl}

\usepackage{graphicx}	% Including figure files
\usepackage{amsmath}	% Advanced maths commands
\usepackage{amssymb}	% Extra maths symbols
\usepackage{multicol}
\usepackage[section]{placeins}
\usepackage{subcaption}
\usepackage{float}
\usepackage{newtxtext,newtxmath}
\title[Nitrogen abundances in SDSS and KLEVER]{The KLEVER survey: Nitrogen abundances at $z\sim$2 and probing the existence of a fundamental nitrogen relation}

\author[C. Hayden-Pawson et al.]{Connor Hayden-Pawson,$^{1,2}$\thanks{E-mail: cjh233@cam.ac.uk}
Mirko Curti,$^{1,2}$
Roberto Maiolino,$^{1,2,3}$
Michele Cirasuolo,$^{4}$\newauthor
Francesco Belfiore,$^{5}$
Michele Cappellari,$^{6}$
Alice Concas,$^{1,2,4,5,7}$
Giovanni Cresci,$^{5}$\newauthor
Fergus Cullen,$^{8}$
Chiaki Kobayashi,$^{9}$
Filippo Mannucci,$^{5}$
Alessandro Marconi,$^{5,7}$\newauthor
Massimo Meneghetti,$^{10}$
Amata Mercurio,$^{11}$
Yingjie Peng,$^{12}$
Mark Swinbank$^{13}$
and \newauthor
Fiorenzo Vincenzo$^{14,15}$
\\\\
$^{1}$ Cavendish Laboratory, University of Cambridge, 19 J. J. Thomson Ave., Cambridge CB3 0HE, UK\\
$^{2}$ Kavli Institute for Cosmology, University of Cambridge, Madingley Road, Cambridge CB3 0HA, UK\\
$^{3}$ Department of Physics and Astronomy, University College London, Gower Street, London WC1E 6BT, UK \\
$^{4}$ European Southern Observatory, Karl-Schwarzschild-Strasse 2, D-85748 Garching bei Muenchen, Germany\\
$^{5}$ INAF - Osservatorio Astrofisico di Arcetri, Largo E. Fermi 5, 50125, Firenze, Italy\\
$^{6}$ Sub-Department of Astrophysics, Department of Physics, University of Oxford, Denys Wilkinson Building, Keble Road, Oxford, OX1 3RH, UK\\
$^{7}$ Dipartimento di Fisica e Astronomia, Universitá di Firenze, Via G. Sansone 1, 50019, Sesto Fiorentino (Firenze), Italy\\
$^{8}$ SUPA Scottish Universities Physics Alliance, Institute for Astronomy, University of Edinburgh, Royal Observatory, Edinburgh EH9 3HJ, UK\\
$^{9}$ Centre for Astrophysics Research, Department of Physics, Astronomy and Mathematics, University of Hertfordshire, Hatfield, AL10 9AB, UK\\
$^{10}$ INAF - Osservatorio di Astrofisica e Scienza dello Spazio di Bologna, via Gobetti 93/3, I-40129 Bologna, Italy\\
$^{11}$ INAF-Osservatorio Astronomico di Capodimonte, Via Moiariello 16, 80131 Napoli, Italy\\
$^{12}$ Kavli Institute for Astronomy and Astrophysics, Peking University, 5 Yiheyuan Road, Beijing 100871, China\\
$^{13}$ Centre for Extragalactic Astronomy, Department of Physics, Durham University, South Road, Durham DH1 3LE, UK\\
$^{14}$ Department of Astronomy \& Center for Cosmology and AstroParticle Physics, The Ohio State University, Columbus, OH 43210, USA\\
$^{15}$ E.A. Milne Centre for Astrophysics, University of Hull, Cottingham Rd, Kingston upon Hull, HU6 7RX, UK
}

\date{Accepted XXX. Received YYY; in original form ZZZ}

\pubyear{2021}

\begin{document}
\label{firstpage}
\pagerange{\pageref{firstpage}--\pageref{lastpage}}
\maketitle

% Abstract of the paper
\begin{abstract}
We present a comparison of the nitrogen-to-oxygen ratio (N/O) in 37 high-redshift galaxies at $z\sim$2 taken from the KMOS Lensed Emission Lines and VElocity Review (KLEVER) Survey with a comparison sample of local galaxies, taken from the Sloan Digital Sky Survey (SDSS). The KLEVER sample shows only a mild enrichment in N/O of $+$0.1 dex when compared to local galaxies at a given gas-phase metallicity (O/H), but shows a depletion in N/O of $-$0.35 dex when compared at a fixed stellar mass (M$_*$). We find a strong anti-correlation in local galaxies between N/O and SFR in the M$_*$-N/O plane, similar to the anti-correlation between O/H and SFR found in the mass-metallicity relation (MZR). We use this anti-correlation to construct a fundamental nitrogen relation (FNR), analogous to the fundamental metallicity relation (FMR). We find that KLEVER galaxies are consistent with both the FMR and the FNR. This suggests that the depletion of N/O in high-$z$ galaxies when considered at a fixed M$_*$ is driven by the redshift-evolution of the mass-metallicity relation in combination with a near redshift-invariant N/O-O/H relation. Furthermore, the existence of an fundamental nitrogen relation suggests that the mechanisms governing the fundamental metallicity relation must be probed by not only O/H, but also N/O, suggesting pure-pristine gas inflows are not the primary driver of the FMR, and other properties such as variations in galaxy age and star formation efficiency must be important. 
\end{abstract}

% Select between one and six entries from the list of approved keywords.
% Don't make up new ones.

%%%%%%%%%%%%%%%%%%%%%%%%%%%%%%%%%%%%%%%%%%%%%%%%%%

%%%%%%%%%%%%%%%%% BODY OF PAPER %%%%%%%%%%%%%%%%%%

\section{Introduction} \label{Intro}

% Intro

The different chemical elements found in the interstellar medium (ISM) of galaxies are produced by different stars and released on different timescales. As such, the relative abundances of these metals can provide compelling diagnostics for investigating many of the processes that drive galaxy evolution. The gas-phase metallicity for example, traced through the oxygen abundance (O/H), is sensitive to the chemical enrichment history of the galaxy through past and current star formation as well as gas inflows and outflows driven by phenomena such as pristine gas accretion and supernovae feedback. The scaling relation between the gas-phase metallicity and the stellar mass (M$_{*}$), known as the mass-metallicity relation (MZR), has has been investigated extensively in literature \citep[see][for a review]{Maiolino_2019}, with more massive systems found to be more chemically enriched. This can be interpreted as massive galaxies representing a more advanced stage of chemical evolution, or as a consequence of massive systems retaining more metals due to their deeper gravitational potentials (\citealt{Kobayashi_2007}; \citealt{SomervilleDave_2015}). As well as being assessed in the local Universe, the mass-metallicity relation has been observed to be already in place in galaxies out to $z\sim$3 (\citealt{Erb_2006}; \citealt{Maiolino_2008}; \citealt{Troncoso_2014}; \citealt{Mannucci_2009}; \citealt{Sanders_2018}), although its normalisation is found to decrease with redshift, as a possible consequence of the combined effect of more prominent gas inflows and stronger gas outflows (\citealt{Sanders_2020FMR}). The scatter in the MZR has also been shown to be partly ascribed to a secondary dependence of the metallicity on the star formation rate (SFR), with several authors finding that the relationship between M$_{*}$, O/H and SFR, referred to as the Fundamental Metallicity Relation (FMR), is redshift invariant out to $z\sim$3 (\citealt{Mannucci_2010}; \citealt{AndrewsMartini_2013}; \citealt{Cresci_2019}; \citealt{Curti_2020FMR}; \citealt{Sanders_2020FMR}). 

Another useful chemical abundance diagnostic is the nitrogen abundance, as probed by the nitrogen-to-oxygen ratio (N/O). Oxygen is an alpha-element, meaning it is mostly produced by He fusion in massive stars, which die as core-collapse supernovae (SNe). In contrast, the production mechanisms for nitrogen are much more complex, making the N/O ratio sensitive to the chemical evolution history of a galaxy. This is mainly because nitrogen can behave as both a primary nucleosynthetic product, i.e with a yield independent of metallicity, and a secondary nucleosynthetic product, whereby the nitrogen yield of the star is dependent on the metallicity of the gas cloud in which it formed. The compelling nature of the N/O ratio means that it has been extensively studied for nearby galaxies and \textsc{Hii} regions by several authors (\citealt{Shields_1991}; \citealt{VilaCostas_1993}; \citealt{vanZee_1998}; \citealt{HenryWorthy_1999}; \citealt{PMC_2009}; \citealt{Perez-Montero_2013}; \citealt{Pilyugin_2012}; \citealt{AndrewsMartini_2013}). Though these studies tend to differ quantitatively depending on the choice of calibrations for deriving chemical abundances, they all agree qualitatively on the nature of N/O in relation to O/H. At low metallicities, N/O tends not vary with O/H, and instead maintains a constant value around log(N/O) $\simeq$ $-$1.5, whilst towards higher metallicities the N/O ratio begins to increase rapidly with O/H.

% explain what causes these trends, massive stars, LIMS, winds, failed SNe

The massive stars that produce the majority of oxygen in the ISM also produce nitrogen, mainly as a secondary element through the Carbon-Nitrogen-Oxygen (CNO) cycle, where nitrogen is formed at the expense of the carbon and oxygen that is already within the star. However, the secondary production of nitrogen from massive stars is not sufficient to reproduce the observed N/O ratios at low metallicities (\citealt{Matteucci_1986}; \citealt{Chiappini_2005}). Additional primary production within massive stars is possible with stellar rotation, which diffuses C and O produced within the star to the outer CNO burning zone, where synthesis of primary nitrogen can occur (\citealt{Chiappini_2006}, \citeyear{Chiappini_2008}). Whilst the addition of stellar rotation can reproduce a low metallicity plateau in N/O (\citealt{Kobayashi_2011}), it is not sufficient for massive stars to reproduce the observed normalisation of the N/O plateau. In order to do so, low and intermediate mass stars (LIMS) must be considered (\citealt{V_K_2018}). Low and intermediate mass stars can produce primary nitrogen when hot-bottom burning occurs in conjunction with the third dredge-up during the asymptotic giant branch phase (see \citealt{Vincenzo_2016}) and references therein). During this phase, the surface of the star becomes enriched with freshly made carbon, from which primary nitrogen can be produced via the CNO cycle. At low metallicities, almost all nitrogen production from LIMS is primary, but at super-solar metallicities the majority of nitrogen produced by LIMS comes from secondary nitrogen production, with yields that increase as a function of the initial stellar metallicity. Overall, chemical evolution models suggest that up to 74\% of nitrogen is produced by LIMS (\citealt{Kobayashi_2020}). Since LIMS typically have longer lifetimes than more massive stars, the nitrogen enrichment of the ISM is delayed compared to that of oxygen, meaning the N/O ratio is sensitive to chemical enrichment timescales. Different mechanisms have also been proposed to explain the increase in N/O with O/H towards greater metallicities. \cite{Vincenzo_2016} suggested differential galactic winds that preferentially remove oxygen in order to reproduce the observed trends (see also \citealt{Magrini_2018}), whilst \cite{V_K_2018a} invoked failed SNe to suppress oxygen yields in massive stars at high metallicity.

Beyond metallicity, the N/O ratio has also been found to positively correlate with stellar mass (\citealt{PMC_2009}; \citealt{Perez-Montero_2013}; \citealt{AndrewsMartini_2013}), which may be a consequence of higher mass galaxies being more evolved, hence having higher metallicities and nitrogen abundances. Just as the mass-metallicity relation was found to have a secondary dependence on SFR, authors such as \cite{Perez-Montero_2013} have studied the variation of star formation rate within the M$_{*}$-N/O plane. However, unlike for the mass-metallicity relation, they find no evidence of a strong secondary dependence on the SFR. This result is interesting within itself as whilst O/H is sensitive to the inflow of pristine/metal-poor gas onto a galaxy, diluting the metallicity, N/O is insensitive to this phenomenon (\citealt{Koppen_2005}; \citealt{Belfiore_2015}; \citealt{Kashino_2016}). If confirmed, the lack of SFR dependency in the stellar mass-N/O plane would suggest that the fundamental metallicity relation is driven mainly by infalls of pristine gas, as opposed to other processes that N/O is also sensitive to, such as the age and star formation history of the galaxy. 

%Mention outflows in the above paragraph. 

Recently, with the advent of many near-infrared telescopes, we have been able to probe chemical abundances in high-redshift galaxies through the use of strong-line diagnostics. It has become clear that high-redshift galaxies have different physical properties to those found locally (e.g., \citealt{Kewley_2013}; \citealt{Steidel_2014}, \citeyear{Steidel_2016}; \citealt{Shapley_2015}, \citeyear{Shapley_2019}; \citealt{Sanders_2016}, \citeyear{Sanders_2020OH}; \citealt{Kashino_2017}; \citealt{Strom_2017}, \citeyear{Strom_2018}) and can be characterised by higher electron densities, harder ionising radiation fields, and highly super solar oxygen-to-iron ratios that are expected for gas with enrichment that is dominated by the products of core-collapse supernovae (\citealt{Steidel_2016}; \citealt{Topping_2020}, \citeyear{Topping_2020b}; \citealt{Cullen_2021}). Several works have also highlighted an increased N/O ratio compared to local galaxies at a fixed metallicity (\citealt{Masters_2014}; \citealt{Shapley_2015}; \citealt{Strom_2017}). Potential explanations for such enhanced nitrogen abundances include an increased population of Wolf-Rayet stars (\citealt{Masters_2014}), or an excess of inflowing pristine gas diluting the metallicity (\citealt{AndrewsMartini_2013}). Similarly, \cite{Masters_2016} suggested that the enhanced N/O values at fixed O/H could be driven by the redshift-evolution of the mass-metallicity relation, with high-$z$ galaxies having a lower metallicities than local galaxies when compared at a fixed stellar mass. Such enhancement would rely on any evolution of the M$_{*}$-N/O relation being slower than that of the mass-metallicity relation. This is supported observationally by papers that use the [\ion{N}{II}]/[\ion{S}{II}] ratio as a proxy for N/O (e.g. \citealt{Perez-Montero_2013}, \citealt{Kashino_2017}), where only a mild decrease in N/O at a fixed M$_{*}$ is observed for high-$z$ galaxies. However studies such as \cite{Strom_2017} that use the [\ion{N}{II}]/[\ion{O}{II}] ratio to determine N/O find a larger decrease in N/O at a fixed stellar mass, suggesting a stronger evolution in the M$_{*}$-N/O relation more similar to that of the mass-metallicity relation. If the M$_{*}$-N/O and M$_{*}$-O/H relations do indeed evolve at a similar pace, then it would suggest that the N/O-O/H relation is redshift-invariant, as supported by simulations, as in \cite{V_K_2018}. 

In this work we review the scaling relations between N/O and other galaxy properties such as O/H, M$_{*}$ and SFR, investigating their evolution with redshift (if any) with the ultimate goal of shedding light on the galaxy evolutionary processes that drive such relations. We do so by using both Sloan Digital Sky Survey (SDSS; \citealt{Abazajian_2009}) optical spectroscopy for local galaxies and data from the Kmos LEnsed Velocity and Emission line Review (KLEVER; \citealt{Curti_2020}) near-infrared survey at high-redshift. 

The paper is structured as follows. In Section \ref{all_data} we describe the data, outline our sample selections and explain our methodology for determining galactic properties such as metallicity and star formation rate. Within this section we also introduce for the first time the full sample of galaxies from the KLEVER survey. In Section \ref{Results} we analyse the scaling relations of nitrogen abundance with a variety of galactic properties for both local and high-redshift galaxies, before summarising our main results in Section \ref{Conc}.  Throughout, we assume a standard $\Lambda$CDM cosmology with $H_0 = 70$ km s$^{-1}$ Mpc$^{-1}$, $\Omega_m$=0.3, and $\Omega_{\Lambda}$=0.7. Throughout the paper, the term metallicity refers to gas-phase oxygen abundance, unless otherwise stated.

\section{DATA AND DERIVED QUANTITIES}\label{all_data}

\subsection{SDSS sample ($z\sim$0)}\label{SDSS_into}

To study local galaxies we draw data from the seventh data release (DR7) of the Sloan Digital Sky Survey (\citealt{Abazajian_2009}). The emission line fluxes and stellar mass estimates for galaxies in this data release are obtained from the MPA/JHU catalogue\footnote{\url{http://www.mpa-garching.mpg.de/SDSS/DR7/}}. 

We required all galaxies to be star-forming, as determined by their position on the [\ion{N}{II}]-BPT diagram, adopting the classification scheme from \cite{Kauffmann_2003b} to remove potential AGN. Following \cite{Curti_2020FMR}, the photometric flags from the DR7 sample were cross-matched with those from the DR12 sample to remove galaxies that were poorly deblended, as well as those with mass correction factors lower than 1, i.e. where the stellar mass derived from the total photometry is lower than the stellar mass derived from the photometry within the fibre. To ensure the [\ion{O}{II}]$\lambda\lambda$3727,29 emission line doublet falls within the wavelength coverage of the SDSS spectrograph, a minimum redshift threshold of $z >$ 0.03 was imposed. This allows us to exploit the same suite of rest-frame optical emission lines in both the local and high-redshift galaxy samples presented in this paper. Following \cite{Mannucci_2010} we adopt a high threshold on the signal-to-noise ratio (SNR) of the H$\alpha$ emission line only (SNR>15) in order to reduce any potential biases in the determination of the metallicities and nitrogen abundances that may be caused by imposing SNR thresholds on weaker optical emission lines such as [\ion{N}{II}]. Taken together, the above criteria give a sample of 127,322 galaxies. However, a small number of galaxies in this sample were found to have anomalously low nitrogen abundances as calculated using the [\ion{O}{II}]$\lambda\lambda$3727,29 emission line doublet. Imposing a threshold of SNR>2 on the [\ion{O}{II}]$\lambda\lambda$3727,29 emission line doublet is sufficient to remove these spurious galaxies, creating a final analysed sample of 126,385 local star forming galaxies. 

The stellar masses for our sample are provided by the MPA/JHU catalogue and have been derived from fits to the photometry, with estimates following \cite{Kauffmann_2003a} and aperture corrections to the total stellar mass following the prescription of \cite{Salim_2007}. All mass estimates were renormalized to a \cite{Chabrier_2003} IMF. The SDSS sample spans a range of  8 $<$ log(M$_*$/M$_{\odot}$) $<$ 11.5 in stellar mass, with a median value of log(M$_*$/M$_{\odot}$) $=$ 10. In order to derive quantities such as metallicity, star formation rate and nitrogen abundance, the optical emission line fluxes must first be corrected for reddening. The Balmer decrement was used to calculate E(B-V)$_{gas}$, assuming case B recombination (H$\alpha$/H$\beta$=2.87) and a \cite{Calzetti_2000} extinction curve. We note that using instead the attenuation curve from \cite{Cardelli_1989} does not change the quantitative or qualitative results of our analysis.

\subsection{The KLEVER survey  ($z\sim$2.2)}

\subsubsection{Sample description and observations}
KLEVER is an ESO Large Programme (197.A-0717, PI: Michele Cirasuolo) that has observed 192 galaxies in the redshift range 1.2 $< z <$ 2.5 using the K-band Multi Object Spectrograph (KMOS) on the VLT (\citealt{Sharples_2013}). KMOS is a near-IR multi-object spectrograph consisting of 24 integral-field units (IFUs) that can operate simultaneously within a 7.2 arcminute circular field. Each individual IFU has a 2.4\arcsec $\times$ 2.4\arcsec field of view with a uniform spatial sampling of 0.2\arcsec $\times$ 0.2\arcsec. The 24 IFUs are fed into three separate spectrographs and detectors, with 8 IFUs assigned per detector. By observing in the $YJ$, $H$ and $K$ band filters (spanning the wavelength ranges 1.025 – 1.344 $\mu$m, 1.456 – 1.846 $\mu$m and 1.92 – 2.46 $\mu$m respectively), KLEVER provides full coverage of the portion of the near-infrared spectrum that is observable from the ground, allowing us to  detect a wide range of strong rest-frame optical nebular emission lines on both a galaxy-integrated and spatially resolved scale. In this paper, we focus only on the galaxy-integrated properties of the KLEVER sample, with the spatially resolved capabilities to be exploited in a future work.

KLEVER observations were taken during ESO observing periods P97-P102, with the full sample consisting of both lensed and un-lensed galaxies. The un-lensed galaxies are observed within the southern CANDELS (\citealt{Grogin_2011}; \citealt{Koekemoer_2011}) fields COSMOS, GOODS-S and UDS. A total of 132 bright star-forming galaxies in these fields were selected at the appropriate redshift using archival data from the KMOS$^{3D}$ survey (\citealt{Wisnioski_2015}). The majority of the un-lensed targets lie in the redshift range $z$ $\in$ [2, 2.6] where deep $K$-band observations (typically a minimum of 8 hours on source) from the KMOS$^{3D}$  survey enable robust detections of the H$\alpha$ and [\ion{N}{II}]$\lambda$6584 emission lines. These observations were then supplemented with additional KLEVER observations in the $YJ$ and $H$ bands, with a typical integration time of 4-6 hours on source for each band. For those sources for which supplementary observations were made in both the $YJ$ and $H$ bands we can obtain detections of rest-frame optical emission lines from [\ion{O}{II}]$\lambda\lambda$3727,29 up to  [\ion{S}{II}]$\lambda\lambda$6716,31. Full details of how many galaxies were observed in each band can be found in Table \ref{tab:KLEVER_obs}. In order to ensure all data within the KLEVER sample has been reduced using the same procedures, the raw $K$-band observations from KMOS$^{3D}$ were re-reduced following the procedure outlined in Section~\ref{DataReduc}.

\begin{table*}
 \centering
 \begin{tabular}{cccccc}
  \hline
  \hline
  Type & Redshift Range & Number of galaxies & YJ Band & H Band & K Band\\
  \hline
  %\textbf{CANDELS}\\
  \textbf{Un-lensed} & [2, 2.6] & 98 & [\ion{O}{II}] & [\ion{O}{III}] + H$\beta$ & H$\alpha$ + [\ion{N}{II}] + [\ion{S}{II}]\\
  & \texttt{"} & 11  & - & [\ion{O}{III}] + H$\beta$ &  H$\alpha$ + [\ion{N}{II}] + [\ion{S}{II}]\\
  & \texttt{"} & 5  & [\ion{O}{II}] & - & H$\alpha$ + [\ion{N}{II}] + [\ion{S}{II}]\\
  & [1.1, 1.7] & 9 &  [\ion{O}{III}] + H$\beta$ & H$\alpha$ + [\ion{N}{II}] + [\ion{S}{II}] & - \\
  & \texttt{"} & 9  &  - & H$\alpha$ + [\ion{N}{II}] + [\ion{S}{II}] & - \\
  \hline
  \textbf{Lensed} & [2, 3.45] & 27 &  [\ion{O}{II}] & [\ion{O}{III}] + H$\beta$ & H$\alpha$ + [\ion{N}{II}] + [\ion{S}{II}]\\
  & [1.1, 1.7] & 33 &   [\ion{O}{III}] + H$\beta$ & H$\alpha$ + [\ion{N}{II}] + [\ion{S}{II}] & [\ion{S}{III}]\\

  \hline
  \hline
 \end{tabular}
  \caption{A breakdown of the observations for the full KLEVER sample. The total sample is a combination of 132 un-lensed and 60 lensed objects. The table shows which rest-frame optical emission lines are detectable in each band for objects within a given redshift range. A ‘-’ denotes that no observations were taken in a given band.}
  \label{tab:KLEVER_obs}
\end{table*}

The remaining galaxies in the KLEVER sample are lensed sources that lie within well studied cluster fields from the CLASH (\citealt{Postman_2012}) and Frontier Fields (\citealt{Lotz_2017}) programs. The main advantage of targeting lensed galaxies is the enhanced spatial resolution and flux sensitivity that allows us to probe high-redshift galaxies with lower masses and star formation rates. The combination of lensed and un-lensed galaxies means galaxies in the total KLEVER sample span a broad range of stellar masses (10$^8$ < M$_*$/M$_{\sun}$ < 10$^{11.5}$) and star formation rates (5 < SFR [M$_{\sun}$/yr] < 300), sampling galaxies on, above and below the main sequence of star formation at $z\sim2$ (\citealt{Noeske_2007}). Due to the enhanced flux sensitivity, shorter integration times were required for the lensed galaxies than the un-lensed, with typical on source exposure times of 11 hours in total (5 hours in the $K$ band and 3 hours each in both the $H$ and $YJ$ bands). We targeted lensed galaxies behind the clusters RXJ2248.7-4431 (also known as AS1063), MACSJ0416.1-2403 and MACS J1149.5+2223. Targets were selected using spectroscopic redshift estimates from a range of sources including the CLASH-VLT survey (\citealt{Rosati_2014}, \citealt{Balestra_2016}), the Grism Lens-Amplified Survey from Space (GLASS, \citealt{Treu_2015}) and VLT/MUSE spectroscopic observations of MACS J1149.5+2223 (\citealt{Grillo_2016}). As detailed in Table \ref{tab:KLEVER_obs}, half of the lensed targets lie in the redshift range $z$ $\in$ [2, 3.45], where detections from [\ion{O}{II}]$\lambda\lambda$3727,29 up to  [\ion{S}{II}]$\lambda\lambda$6716,31 allows a direct comparison to the majority of the un-lensed KLEVER objects. However, the other half of the lensed targets fall in the redshift range $z$ $\in$ [1.1, 1.7], for which we can no longer detect the [\ion{O}{II}]$\lambda\lambda$3727,29 doublet in the $YJ$ band, and instead gain access to the [\ion{S}{III}]$\lambda\lambda$9068, 9530 doublet in the K band.

For all of our observations we adopted a object-sky-object (OSO) nodding strategy, with 300s exposures and dithering for sky sampling and subtraction. Within each pointing a maximum of 21 IFUs were allocated to targeted galaxies, with the remaining 3 IFUs assigned to bright stars (one for each detector within KMOS). The average seeing of the observations was calculated by fitting an elliptical Gaussian to the collapsed images of the stars, and was typically found to be 0.5$\arcsec$ - 0.6 $\arcsec$ (FWHM) for our observations. We further used the observations of the bright stars to align exposures and validate the flux calibration, which is discussed in Section \ref{DataReduc}.

\subsubsection{Stellar masses}

For the un-lensed galaxies, derived parameters such as stellar mass were accessed through the KMOS$^{3D}$ data release\footnote{\url{https://www.mpe.mpg.de/ir/KMOS3D/data}} (\citealt{Wisnioski_2019}). The stellar mass estimates for these galaxies were derived following \cite{Wuyts_2011}, wherein the FAST (\citealt{Kriek_2009}) code was used to fit the optical to 8 $\mu$m SEDs with \cite{Bruzual_2003} (BC03) models assuming solar metallicity, a \cite{Calzetti_2000} reddening law, and either constant or exponentially declining star formation histories. 

For the lensed galaxies, stellar masses were estimated following a similar procedure to that adopted in \cite{Curti_2020}. We used the high-$z$ extension of the \textsc{MAGPHYS} code (\citealt{deCunha_2015}) to fit the SEDs for our galaxies, using data from deep Frontier Field \emph{HST} images\footnote{\url{https://irsa.ipac.caltech.edu/data/SPITZER/Frontier/}} (\citealt{Lotz_2017}) in three optical bands (F435W, F606W, F814W) and four near-IR bands (F105W, F125W, F140W, F160W), plus additional  3.6$\mu$m and 4.5$\mu$m IRAC data acquired by Spitzer (\citealt{Castellano_2016}; \citealt{DiCriscienzo_2017}; \citealt{Bradac_2019}). MAGPHYS uses BC03 stellar population synthesis and adopts the two-component model of \cite{Charlot_2000} to describe the attenuation of stellar emission at  ultraviolet, optical, and near-infrared wavelengths. A \cite{Chabrier_2003} IMF is assumed throughout, with the final estimates for the lensed galaxies also being corrected for magnification by considering the median magnification factor for each source taken from the multiple publicly available lensing models of the Frontier Fields\footnote{\url{https:archive.stsci.edu/prepds/frontier/lensmodels/}} (e.g. \citealt{Zitrin_2013}; \citealt{Caminha_2017}; \citealt{Jauzac_2016}...). The methods used to build the public models of the Frontier Fields have been tested and validated with image simulations by \cite{Meneghetti_2017}. We investigated the systematic differences in stellar mass estimates between the lensed and un-lensed galaxy samples by calculating stellar masses for a sub-sample of GOODS-S galaxies using the method applied to our lensed galaxy sample. We found that our method tend to underestimate stellar masses by an average of 0.1 dex compared to KMOS$^{3D}$ estimates.

\subsubsection{Data Reduction} \label{DataReduc}

The raw KMOS observations were reduced using the pipeline provided by ESO (v2.6.6) . Within the pipeline, the advanced sky subtraction method SKYTWEAK from \cite{Davies_2007} was implemented to reduce sky line residuals. Over the course of each observing block (OB), the telescope can drift from its initial position, causing shifts in the position of the target between different exposures. To detect these shifts, we measured the centroid positions of the observed reference stars within each exposure and calculated the relative shifts between them. These shifts were then applied to all scientific sources that were observed on the same detector as the corresponding star. This methodology was applied for combining both individual exposures within a single OB, as well as when combining observations between different OBs. The final datacubes were then created using a three-sigma clipped average over all exposures, excluding those with large seeing values (>0.8$\arcsec$) or unusually large shifts in the measured centroid position of the corresponding star. These datacubes contain 2 main extensions: the first containing the measured flux, and the second containing the associated noise. Finally, the datacubes were reconstructed onto a 0.1$\arcsec$ $\times$ 0.1$\arcsec$ pixel scale.

Where possible, the accuracy of the flux calibration for the final datacubes was checked using the bright stars. In each filter, an image of each star was produced by collapsing over the wavelength axis in order to observe the continuum emission. An aperture of radius 0.8$\arcsec$ was centred on the image, and the total flux within the aperture was extracted. This total flux estimate was used to cross-check the estimated magnitude of the star against the corresponding magnitude provided by the 2MASS stellar catalogue (\citealt{Skrutskie_2006}). The total flux estimates from the KLEVER observations were typically within three sigma of the magnitude estimates and errors provided by 2MASS, suggesting a good degree of accuracy in the flux calibration for the full KLEVER sample.

%The stellar magnitude estimates from 2MASS were found to be in good agreement with those derived from the KLEVER observations in each band,  suggesting a good degree of accuracy in the flux calibration for the full KLEVER sample.

\subsubsection{Emission line fitting}
\label{Emission_line_fit}

To measure the global properties of the KLEVER galaxies, galaxy-integrated spectra were extracted from the reduced datacubes using a circular aperture with a diameter of 1.3\arcsec. The choice of size for the aperture is arbitrary. A larger aperture will ensure more of the galaxy is sampled spatially, making the integrated-flux estimates more representative of the total flux within the galaxy, however it will also increase the noise introduced to the spectra, potentially making it harder to detect faint emission lines. Our choice of size, sampling a physical scale of $\sim$11 kpc from $z$ $\in$ [1.4, 2.2], was found to be the best compromise between these two factors. In each band, we created a continuum image of each galaxy by collapsing over the full wavelength range whilst masking strong OH emission lines as identified using the \cite{Rousselot_2000} catalogue. The centre of the continuum emission in all images was determined through fitting with an elliptical Gaussian profile, and the pseudo-aperture was centred on this location. In cases where no continuum was detected, an image was made by collapsing around the strongest emission line within that band (typically [\ion{O}{II}], [\ion{O}{III}], H$\alpha$ or [\ion{S}{III}]), with the centre being determined using the same procedure as earlier. Examples of the aperture placement for a galaxy where the continuum is detected and another for which there is no detected continuum are given in Appendix~\ref{sec:KLEVER_spec}. If neither continuum nor strong emission line images could be created, then the aperture was placed at the centre of the datacube. The aperture was then used to extract two spectra from each datacube: an integrated flux spectrum from the first extension, and an associated noise spectrum from the second extension.

The emission line features of the spectra were fit using the least-square minimisation method from the python routine LMFIT (\citealt{Newville_2016}). A straight line was used to define the local continuum, with the slope and normalisation allowed to vary as free parameters. Fitting of a proper stellar continuum is not required as the observed continuum is too faint and the equivalent widths of the detected emission lines are large. All emission lines within a given band were simultaneously fit with single Gaussian profile components that were linked in velocity and line width. The tying of the velocity and line widths is necessary to improve the fit to the data, and is a reasonable assumption to make as the emission is expected to originate from similar nebular regions and with similar kinematics. However, velocities and line widths between different observing bands were not tied due to the different resolving powers within KMOS (3582, 4045 and 4227 in the centre of the $YJ$, $H$ and $K$ bands, respectively). The amplitudes of several emission lines were also constrained to physical values. The [\ion{O}{II}] and [\ion{S}{II}] emission line doublets both have ratios that are dependent on the density of the gas, and were allowed to vary between 0.35 < [\ion{O}{II}]$\lambda$3729/[\ion{O}{II}]$\lambda$3727 < 1.50 and 0.44 < [\ion{S}{II}]$\lambda$6716/[\ion{S}{II}]$\lambda$6731 < 1.45, respectively (\citealt{Osterbrock_2006}). Similarly, the [\ion{N}{II}]$\lambda$6584/[\ion{N}{II}]$\lambda$6548 and [\ion{O}{III}]$\lambda$5007/[\ion{O}{III}]$\lambda$4958 emission line ratios are both constrained to values of 3 by atomic physics. Each fit was inverse weighted by its corresponding noise spectrum. To minimise contamination from OH emission lines, the spectra were entirely masked at the positions of the brightest of these lines, as identified using the catalogue presented in \cite{Rousselot_2000}. Examples of the best-fit models to the integrated spectra of galaxies in the KLEVER sample are shown in Appendix~\ref{sec:KLEVER_spec}. The noise near each emission line was estimated by taking the root mean square of the flux in two continuum regions either side of the emission line. To determine errors on the flux, velocity and line width values, we repeated the fit 100 times, perturbing the flux values in each wavelength channel by a random number drawn from a Gaussian profile distribution, the width of which was defined by the noise value in the corresponding channel of the noise extension within the datacube. A 2.5-sigma clipping was applied to all of the resulting distributions to remove poor fits to the perturbed data. For each parameter, the estimates for the lower and upper errors were taken to be the 16th and 84th percentiles of the distribution. 

To report an emission line as detected, we set thresholds on both the SNR and the percentage error on the final parameters. We require the strongest line observed in a given band (H$\alpha$, [\ion{O}{III}], [\ion{O}{II}], [\ion{S}{III}]) to have a SNR>3 and a percentage error on the flux of less than 33\%. For the weaker lines within these bands, which are tied to the detection of the strongest line, we reduce the requirements to a SNR>2 and percentage error of less than 50\%. We obtain a detection rate of 88\% for H$\alpha$, 62\% for [\ion{N}{II}]$\lambda$6584, 45\% for [\ion{S}{II}]$\lambda$6717 and 38\% for [\ion{S}{II}]$\lambda$6731. We note that the lower detection rate for the [\ion{S}{II}] emission line doublet is partly driven by these lines falling primarily at the far end of the $K$-band, where rising thermal emission causes increased noise in the near-infrared spectra. [\ion{O}{III}] is detected for 60\% of the sample, and H$\beta$ in 52\%. Within the 129 galaxies for which the [\ion{O}{II}]$\lambda\lambda$3727,29 doublet can be observed, we obtain a detection rate of 72\%, whilst for the galaxies at $z$ $\approx$ 1.4, we detect [\ion{S}{III}]$\lambda$9530 in 45\% of objects. 

The analysis within this paper relies on the simultaneous detection of several rest frame optical emission lines needed to derive the metallicity and nitrogen abundance. As such, we include only KLEVER galaxies for which the [\ion{O}{II}], H$\beta$,  [\ion{O}{III}], H$\alpha$ and [\ion{N}{II}] emission lines have all been detected. We remove a further 6 galaxies from our analysis that were identified as potential AGN based off of their position on the [\ion{N}{II}]-BPT diagram using the classification from \cite{Kauffmann_2003b}. As described for the local sample in Section \ref{SDSS_into}, the Balmer decrement was used to correct all emission lines for reddening. To ensure this correction is applicable to the full sample, we remove 3 galaxies that have spuriously large Balmer decrements, potentially caused by strong OH emission lines near the H$\beta$ line corrupting the fit to the data. When any line ratio involves an extinction-corrected emission line (e.g. N2O2), we account for the uncertainty from the correction as follows. We perturb the H$\alpha$ and H$\beta$ fluxes within their respective errors, before using the perturbed Balmer decrement ratio to calculate the extinction corrected fluxes and using them to calculate the appropriate extinction-corrected line ratio. If a perturbed Balmer decrement falls below the intrinsic case B value of 2.87, we do not correct for extinction. We repeat this process 1000 times and take the errors on the extinction-corrected line ratio to be the 16th and 84th percentiles of the resulting distribution. To get a final error estimate for the corrected ratio, these extinction errors are summed in quadrature with the error on the line ratio propagated from the uncorrected line flux measurements. We note that the error from the uncorrected line fluxes is still dominant over the error introduced from the extinction correction.

The final sample of KLEVER galaxies analysed within this paper then consists of 37 galaxies, 6 of which are lensed, spanning a range of 10$^{9.5}$ < M$_*$/M$_{\sun}$ < 10$^{11.5}$ in stellar mass.

\subsection{Star formation rates}\label{sec:SFR}

Star formation rates were measured from the dust-corrected H$\alpha$ emission line luminosity, using the conversion factor between H$\alpha$ and SFR from \cite{Kennicutt_2012}. These were then renormalized to a \cite{Chabrier_2003} IMF in order to align with the stellar mass estimates. For the SDSS sample, the low redshift cut of $z >$ 0.03 means that some galaxies may be included where only the central $\sim1.6$ kpc falls within the SDSS fibre, meaning only the inner-most regions of the galaxy are sampled. As such, estimating the star formation rate from the flux within the fibre may lead to systematic underestimations of the total SFR within the galaxy. To obtain estimates for the total star formation rate for our local galaxies, aperture corrections provided by the MPA/JHU catalogue were applied. These corrections build on the work of \cite{Salim_2007}, where aperture-corrected  H$\alpha$ fluxes from \cite{Brinchmann_2004} were shown to be in good agreement with UV-based SFR estimates, with an average offset of just 0.06 dex and no correlation between the residuals and the level of aperture correction applied. \cite{Richards_2016} also showed that aperture corrected SFR estimates are accurate to within 0.2 dex for large samples of galaxies.

%Although such corrections can introduce uncertainties into the SFR estimation, the resulting star formation rates are more representative of the global star formation occurring within a galaxy.

For the KLEVER galaxies, the fluxes have been extracted from a circular aperture of diameter 1.3$\arcsec$, as discussed in Section \ref{Emission_line_fit}, corresponding to a physical distance of roughly 11 kpc at redshift $\sim 2.2$. As such, the majority of the H$\alpha$ flux is contained within the circular aperture, therefore we apply no aperture corrections to the KLEVER data. Since the SFR estimates for the KLEVER galaxies are more similar to total SFR values, hereafter we present only the results using total SFR estimates for the SDSS sample, allowing a more direct comparison between our high and low redshift samples.

\subsection{Metallicity determination}\label{OH_det}

% table defining some line ratios used commonly within the paper
\begin{table}
 \centering
 \begin{tabular}{lc}
  \hline
  \hline
  Notation & Line Ratio\\
  \hline
  R2 & [\ion{O}{II}]$\lambda\lambda$3727,29 / H$\beta$\\
  R3 & [\ion{O}{III}]$\lambda$5007 / H$\beta$\\
  N2 & [\ion{N}{II}]$\lambda$6584 / H$\alpha$\\
  S2 & [\ion{S}{II}]$\lambda\lambda$6717,31 / H$\alpha$\\
  R23 &( [\ion{O}{II}]$\lambda\lambda$3727,29 + [\ion{O}{III}]$\lambda\lambda$4959, 5007) / H$\beta$\\
  O32 & [\ion{O}{III}]$\lambda$5007 / [\ion{O}{II}]$\lambda\lambda$3727,29\\
  N2O2 & [\ion{N}{II}]$\lambda$6584 / [\ion{O}{II}]$\lambda\lambda$3727,29\\
  N2S2 & [\ion{N}{II}]$\lambda$6584 / [\ion{S}{II}]$\lambda\lambda$6717,31\\
  \hline
  \hline
 \end{tabular}
  \caption{Definitions of line ratios adopted throughout this paper.}
  \label{tab:diag_notations}
\end{table}

Metallicity can be determined using strong-line diagnostics that are calibrated through measurements of the temperature-sensitive ratio of auroral and nebular emission lines. These auroral lines are typically extremely weak and difficult to detect in distant galaxies, meaning many strong-line diagnostics that are currently applied to high-redshift galaxies are calibrated using samples of local galaxies and \textsc{Hii} regions. However, it has been well documented that the \textsc{Hii} regions within high-redshift star-forming galaxies have different physical properties to those found locally (e.g., \citealt{Kewley_2013}; \citealt{Steidel_2014}; \citealt{Shapley_2015}, \citeyear{Shapley_2019}; \citealt{Sanders_2016}; \citealt{Kashino_2017}; \citealt{Strom_2017}), which may mean that these locally-calibrated metallicity diagnostics can yield biased metallicity estimates when applied to high-redshift galaxies. Efforts have been made to calibrate metallicity diagnostics at high redshifts, but detections of the essential auroral lines have been possible for just a handful of galaxies at high-redshift, allowing only preliminary calibrations of strong-line diagnostics to be made (\citealt{Sanders_2016auro}, \citeyear{Sanders_2020OH}). Whilst these results give us an early insight into how the calibrations of certain diagnostics may evolve with redshift, the limited sample size and extreme nature of the galaxies sampled means that they may not be representative of the full galaxy population at high redshift. As such, we must await new facilities like JWST or the MOONS spectrograph on the VLT (\citealt{Cirasuolo_2020}; \citealt{Maiolino_2020}) before we can obtain robust metallicity diagnostics calibrated using direct temperatures at high redshifts. Until then we can only make informed decisions on our choice of metallicity indicators and diagnostics in the knowledge of their own potential biases.

Of particular concern are those diagnostics which include nitrogen. The N2 line ratio is commonly used to determine metallicities in galaxies at high-redshift as the close proximity of the H$\alpha$ and [\ion{N}{II}] lines means that the ratio is insensitive to dust extinction and observations in only one band are required. However, the N2 ratio has been found to overestimate the metallicities of high redshift galaxies (\citealt{Liu_2008}, \citealt{Newman_2014}). N2 is known to correlate with the N/O ratio, as discussed in \cite{PMC_2009}, meaning the discrepancy at high redshifts may be explained by elevated N/O at a fixed O/H (\citealt{Masters_2014}, \citeyear{Masters_2016}; \citealt{Shapley_2015}), whilst an increase in the hardness of the ionising spectrum at a fixed N/O and O/H has also been suggested as a caused of elevated N2 ratios (\citealt{Steidel_2016}; \citealt{Strom_2017}). Indeed, the assumed relation between N/O and O/H is itself important when using strong-line metallicity diagnostics, as those involving nitrogen rely on N/O varying similarly as a function of O/H for both the calibration sample and the sample that the calibration is being applied to. Whilst several studies have found the N/O-O/H relation to be qualitatively similar, with a low-metallicity N/O plateau and an increasing N/O at higher metallicities, a large scatter among individual objects has been observed. Furthermore, systematic differences between the exact parameterisations of this relationship exist depending on the methods and calibrations adopted, as discussed in Section \ref{Intro}. The nature of the N/O-O/H relation at high redshift is also not robustly determined, and although there are suggestions from both observations (\citealt{Strom_2018}) and simulations (\citealt{V_K_2018}) that the relationship may be redshift invariant, the large scatter of N/O at a given O/H observed in high-redshift galaxies could mean that applying local calibrations may not be appropriate. 

To avoid some of these issues, it may be more appropriate to use strong-line calibrations based only on alpha-elements, such as oxygen, at high redshift. Studies such as \cite{Shapley_2015} have shown that high redshift galaxies are not offset from local galaxies in oxygen-based diagnostics diagrams, such as R23  v O32. However diagnostic diagrams such as these may be more sensitive to the ionisation conditions within the galaxy, as opposed to the metallicity (see \citealt{Strom_2018}). Recently, \cite{Sanders_2020OH} compared metallicity diagnostics calibrated on local galaxies, \textsc{Hii} regions and high-$z$ analogues to a sample of high redshift galaxies for which direct temperature metallicity calibrations were possible. They found that the O32 diagnostic, which has the advantage of being monotonic across the range of metallicities probed, has a scatter for high redshift galaxies at a fixed O/H that is similar to that of local galaxy samples. Furthermore, the calibrations based on local galaxy samples from \citeauthor{Curti_2017} (\citeyear{Curti_2017}, \citeyear{Curti_2020FMR}) as well as those from local high-redshift analogues (\citealt{Bian_2018}) are in good agreement with the median values of the high redshift sample for several oxygen-based line ratios (R3, R2, R23, and O32). As such, within this paper we adopt the metallicity diagnostics provided in \cite{Curti_2020FMR}, which have been calibrated for application in the metallicity range 12 + log(O/H) $\in$ [7.6, 8.9]. We use only the R3, R2, R23, and O32 diagnostics, and exclude any diagnostics which make use of nitrogen emission lines. This is particularly important as we derive both metallicity and nitrogen abundance (see Section \ref{NO_det}) using strong-line diagnostics. It is therefore essential that we do not rely on the same nitrogen emission lines for the determination of both of these quantities in order to avoid introducing artificial correlations between them. Combinations of different diagnostics are used simultaneously to constrain the metallicity and break the degeneracies that can occur when using only a single diagnostic. The final metallicities and associated uncertainties were calculated using a simultaneous chi-squared minimisation over all available line ratios, as described in \cite{ Curti_2020}.  

The lack of aperture-correction for the metallicity estimates means that the global metallicity is assumed to be similar to that of the central region of the galaxy probed by the SDSS fibre. Locally, higher mass galaxies tend to have steeper metallicity (as well as N/O) gradients (\citealt{Belfiore_2017}). This could mean that, particularly for higher mass galaxies, we are likely biased towards larger O/H and N/O estimates. However, by comparing fibre-based SFR estimates to our aperture-corrected SFR estimates from Section~\ref{sec:SFR}, we estimate that 75\% of our SDSS sample have fibre covering fractions greater than 20\%, above which abundances from central regions are expected to agree well with integrated values (\citealt{Kewley_2005}).

\subsection{Nitrogen abundances} \label{NO_det}

When comparing metallicities to nitrogen abundances, it is important to ensure that they have both been estimated in a self-consistent method to allow direct comparisons between the two. As such, it is vital that we adopt a strong-line diagnostic for the nitrogen abundance that has been calibrated using the same sample as the metallicity diagnostics used. We therefore adopt a new calibration for N/O based on the `direct', electron-temperature method for abundance determination applied to the sample of stacked SDSS galaxy spectra described in \cite{Curti_2017}. 

This calibration relies primarily on the assumption that, given the similar ionisation potentials of N and O, the total N/O abundance can be estimated from the ratio of the individual ionic abundances, e.g. the relative single-ionised species (N/O $\simeq$ N$^+$/O$^+$, \citealt{Garnett_1990}, \citealt{VilaCostas_1993}). 
In more detail, for each of the stacked spectra in the \cite{Curti_2017} sample (corresponding to different bins in the [\ion{O}{ii}]$\lambda3727,29$/H$\upbeta$ vs [\ion{O}{iii}]$\lambda5007$/H$\upbeta$ diagram) we derive the electron temperature of the `low-ionisation' region (characterised, indeed, by the presence of single-ionised oxygen and nitrogen species) from the auroral-to-nebular [\ion{O}{ii}]$\lambda7320,30$/[\ion{O}{ii}]$\lambda3727,29$ line ratio. We use such temperatures to compute the abundance of O$^{+}$ and N$^{+}$ from the [\ion{O}{ii}]$\lambda3727,29$/H$\upbeta$ and [\ion{N}{ii}]$\lambda6584$/H$\upbeta$ ratio, respectively, exploiting the PyNeb package from \cite{Luridiana_2015}. The electron density is constrained independently, for each spectrum, from the [\ion{S}{ii}]$\lambda6717$/[\ion{S}{ii}]$\lambda6731$
diagnostic ratio.
We note here that we adopt the same electron temperature based on the O$^{+}$ auroral-to-nebular line ratio to derive both O$^{+}$ and N$^{+}$ ionic abundances. 
In fact, although some evidence exists that t$_{2}[\ion{N}{ii}]$ and t$_{2}[\ion{O}{ii}]$ might differ when measured in individual \textsc{Hii} regions (with t$_{2}[\ion{O}{ii}]$ higher, on average, than t$_{2}[\ion{N}{ii}]$, as potentially driven by the contribution to the [\ion{O}{ii}]$\lambda7320,30$ auroral line flux from dielectronic recombination), the two temperatures are found to be consistent within their uncertainties in our stacked spectra (see Fig. 4 of \citealt{Curti_2017}), as also reported in different studies (see e.g., \citealt{Esteban_2009}; \citealt{Pilyugin_2009}; \citealt{Berg_2015},\citeyear{Berg_2020}).
Therefore, we prefer to self-consistently model the low-ionisation zone with only one temperature (we choose t$_{2}[\ion{O}{ii}]$ over t$_{2}[\ion{N}{ii}]$ because it is measured with higher signal-to-noise and in more stacked spectra, and because it has been used to infer the global oxygen abundance as well) rather than introducing systematics by adopting two different temperatures for the two ionic species.
%because of the low S/N detections of the [\ion{N}{ii}]$\lambda5755$ auroral line in many of the analysed spectra; nonetheless, it is reasonable to assume that the temperatures derived from the two different auroral lines are comparable, as both ions probe the same ionisation region.
Finally, the N/O abundance in our stacks is assumed to be traced by the N$^+$/O$^+$ abundance, and this is calibrated against different strong line ratios in a similar fashion as done for the full oxygen abundance calibrations.

\begin{figure}
    \centering
    \includegraphics[width=\columnwidth]{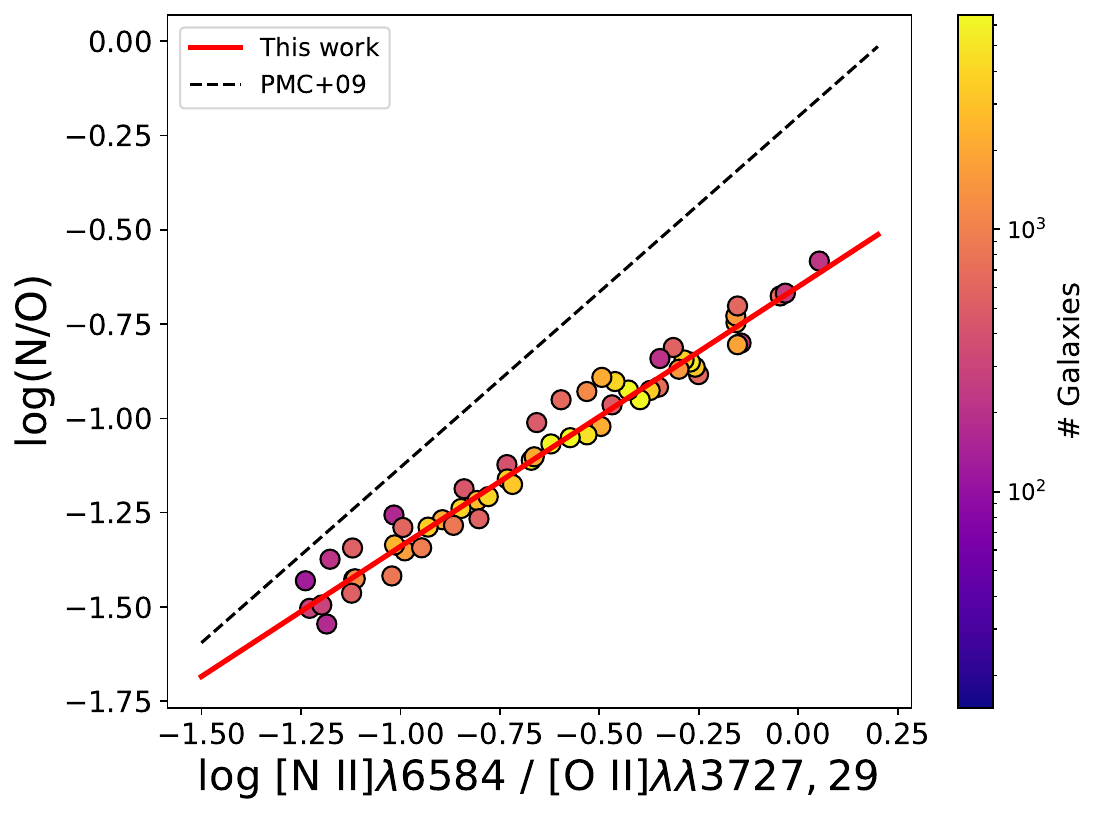}
    \caption{The N/O abundance as a function of N2O2, as measured via detection of auroral lines in stacked spectra of SDSS galaxies (\citealt{Curti_2017}). The red line represents the linear fit to the data for the new N/O calibration provided by Equation~\ref{N2O2_eq}. Each point is colour-coded according to the number of galaxies within each stack, whilst the dashed black line shows, for reference, the log(N2O2) vs log(N/O) calibration presented by PMC09.}
    \label{fig:N2O2_calib}
\end{figure}

Figure \ref{fig:N2O2_calib} shows the relationship between the T$_{e}$-based N/O abundance inferred for each stacked spectrum as a function of the [\ion{N}{II}]/[\ion{O}{II}] line ratio, with color-coding indicating the number of galaxies used in each stacked spectrum. The red line is a linear fit to the data, providing the following calibration
\begin{equation}\label{N2O2_eq}
\text{log(N/O)} = \left(0.69 \times \text{log(N2O2)}\right) - 0.65,
\end{equation}
with the black dashed line showing the calibration from \cite{PMC_2009} (hereafter PMC09) as a reference. Similar to the calibration presented in \cite{Strom_2017} that used a calibration sample of extragalatic \textsc{Hii} regions from \cite{Pilyugin_2012}, we find that the PMC09 calibration may provide N/O estimates that are ~0.3 dex larger than our own calibration at high N2O2 values. The main cause of the discrepancy between the PMC09 calibration and our own is the different calibration samples. The PMC09 sample is mostly made up of \ion{H}{II} galaxies that have low metallicities, with N2O2 values that are typically below -0.75. These galaxies dominate the linear best-fit to the data, and the smaller number of Giant Extragalactic \ion{H}{II} Regions that populate the high N2O2 region of the PMC09 sample tend to fall below this best-fit calibration line. By contrast, our stacked SDSS galaxy sample is homogenous and extends up to N2O2 values greater than 0, providing a best-fit calibration that better represents galaxies with high N/O values. Indeed, the Giant Extragalactic \ion{H}{II} Regions that populate the high N2O2 end of the PMC09 sample are well represented by our own calibration.

\begin{figure}
    \includegraphics[width=\columnwidth]{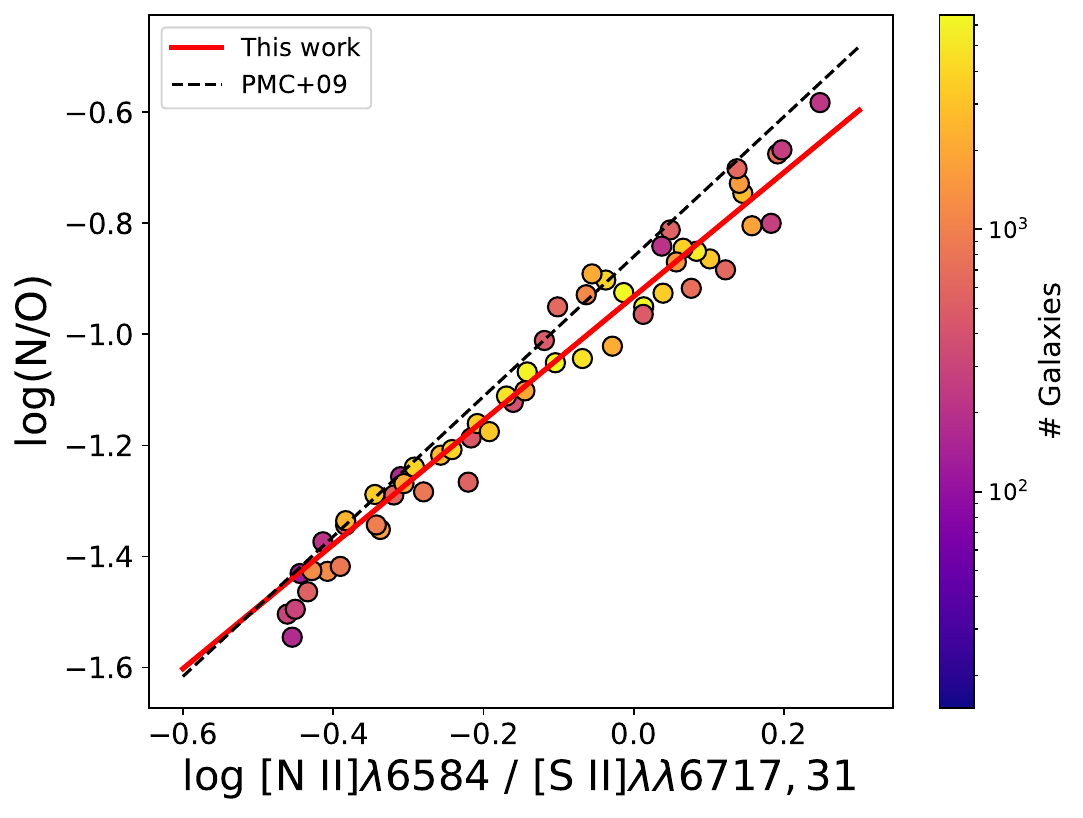}
    \caption{The N/O abundance as a function of N2S2, as measured via detection of auroral lines in stacked spectra of SDSS galaxies (\citealt{Curti_2017}). The red line represents the linear fit to the data for the new N/O calibration provided by Equation \ref{N2S2_eq}. Each point is colour-coded according to the number of galaxies within each stack, whilst the dashed black line shows, for reference, the log(N2S2) vs log(N/O) calibration presented by PMC09.}
    \label{fig:N2S2_calib}
\end{figure}

A number of studies have used the N2S2 diagnostic, first introduced in \cite{Sabbadin_1977}, to estimate the N/O abundance. The clear advantage of this diagnostic over N2O2 is that the proximity of the [\ion{N}{II}] and [\ion{S}{II}] emission lines means no correction for extinction due to dust is required and observations in only a single band are needed to obtain simultaneous detections. Figure \ref{fig:N2S2_calib} shows the relationship between the T$_{e}$-based N/O abundance inferred for each stacked spectrum from \cite{Curti_2017} as a function of the [\ion{N}{II}]/[\ion{S}{II}] line ratio, with colour-coding indicating the number of galaxies used in each stacked spectrum. The red line is a linear fit to the data, providing the following calibration
\begin{equation}\label{N2S2_eq}
\text{log(N/O)} = \left(1.12 \times \text{log(N2S2)}\right) - 0.93.
\end{equation}

\begin{figure}
    \centering
    \includegraphics[width=\linewidth]{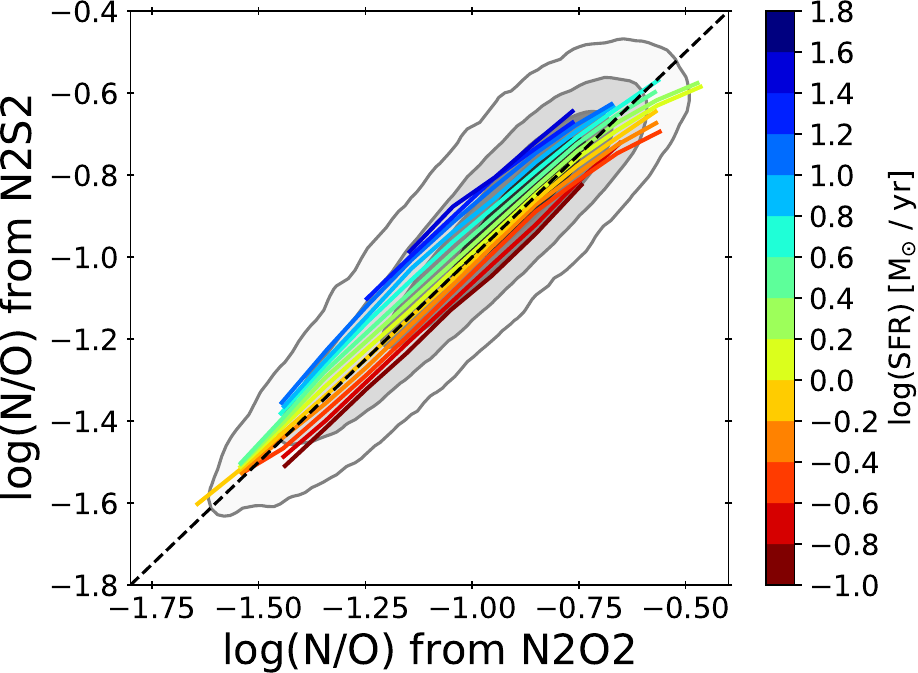} \par
    \caption{N/O derived from both N2O2 and N2S2 for galaxies in our SDSS sample, with contours showing the regions encompassing 30\%, 60\% and 90\% of the total sample. The coloured lines show how the relationship between N2S2 and N2O2 varies as a function of star formation rate. At a fixed N2O2, more star forming galaxies have systematically larger values of N2S2, causing a discrepancy in the estimation of N/O.}
    \label{fig:N2O2vN2S2}
\end{figure}

Whilst oxygen has a relatively similar ionisation potential to nitrogen (13.6 eV and 14.6 eV respectively), the lower ionisation potential of sulphur (10.4 eV) means that the N2S2 diagnostic is more sensitive to the ionisation conditions within the gas than N2O2. To investigate some of the differences that arise when using either N2O2 or N2S2, we estimate the nitrogen abundance from both N2O2 and N2S2, using Equations \ref{N2O2_eq} and \ref{N2S2_eq} respectively, for our full SDSS sample. The results are shown as grey contours in Figure \ref{fig:N2O2vN2S2}. Overlaid on these contours, we plot tracks of constant star formation rate taken in 0.2 dex wide SFR bins. A similar analysis considering specific star formation rate is presented in Appendix \ref{sec:sSFR_appendix}. A clear trend emerges where those galaxies with higher star formation rates have a nitrogen abundance estimated from N2S2 that is systematically greater than that estimated using N2O2. The difference in N/O as estimated using N2S2 at a fixed N2O2 can vary by more than 0.2 dex, depending on the star formation rate.

To better understand the origin of this difference, we must consider what physical properties each line ratio probes, and how those relate to N/O. To first order, N2O2 and N2S2 are strongly correlated as oxygen and sulphur are both $\alpha$ elements, meaning both ratios trace N/$\alpha$. However, differences between these ratios can arise due to the unique chemical enrichment histories and physical properties of both oxygen and sulphur respectively. \cite{Kobayashi_2020} suggest that a considerable fraction of sulphur can be produced by Type 1a supernovae, meaning N2S2 may be a less effective tracer of N/$\alpha$ than N2O2. Furthermore, the abundances of O$^+$ and S$^+$ are strongly dependent on the ionisation conditions within a given galaxy. Models of \textsc{Hii} regions suggest that O$^+$ has a very similar ionisation structure to N$^+$, whilst S$^+$ differs in that it is mostly emitted by the outer shell of the \textsc{Hii} region, with higher ionisation species such as S$^{++}$ dominating the inner regions (\citealt{Levesque_2010}; \citealt{Mannucci_2021}). As the star formation within a galaxy increases, the number of \textsc{Hii} regions within the galaxy will grow, and these regions may merge. As such, the integrated spectrum of the ionised gas within the galaxy (as observed through an aperture in the case of SDSS) will begin to be dominated by more central \textsc{Hii} regions, leading to a decrease in observed S$^+$, and hence an increase in N2S2. Meanwhile the similar ionisation structures of N$^+$ and O$^+$ mean N2O2 is relatively unaffected. This effect could be particularly significant for high-redshift star forming galaxies, which tend to resemble large \textsc{Hii} regions. The low S/N of high-redshift observations means that the outer regions of these galaxies may be undetected, meaning less [\ion{S}{II}] is detected and leading to larger discrepancies between N2S2 and N2O2.

\cite{Strom_2018} found that both high-redshift galaxies and local \textsc{Hii} regions were offset from the trend between N2O2 and N2S2 seen in local galaxies. They proposed that the integrated light spectrum of local galaxies may include contributions from diffuse ionised gas (DIG) in addition to emission from \textsc{Hii} regions. Galaxies that have a larger contribution from DIG would be expected to have enhanced flux contribution to the [\ion{S}{II}] doublet, which would depress N2S2 at a fixed N2O2. However, recent work by \cite{Mannucci_2021} suggests that SDSS galaxies do not have a significant contribution from DIG, and that instead aperture effects on the spectroscopy of nearby \textsc{Hii} regions are largely responsible for the observed differences. Indeed, when large enough apertures are used on local \textsc{Hii} regions, the observed line ratios are the same as in SDSS galaxies.

Due to the caveats discussed above, we caution the use of the N2S2 diagnostic to trace N/O, particularly for high-$z$ galaxies and \textsc{Hii} regions wherein [\ion{S}{II}] emission from the outer ionised regions may not be properly represented in integrated spectra because of low S/N or small extraction apertures, respectively. Furthermore, the differing enrichment histories of sulphur and oxygen make N2O2 a far more direct tracer of N/O than N2S2. These points, coupled with the relatively low detection rate of the full [\ion{S}{II}] doublet for our KLEVER sample, compel us to adopt only the N2O2 diagnostic throughout this paper.

%This trend may be driven by different ionisation conditions at different star formation rates. Galaxies with a lower star formation \textcolor{red}{rate} may preferentially ionise species with lower ionisation potentials, which would increase the fraction of S$^+$ relative to O$^+$ and N$^+$, decreasing N2S2, whilst the more similar ionisation potentials of O$^+$ and N$^+$ would leave N2O2 relatively unaffected.

%This effect is particularly important when considering high-redshift star forming galaxies, which tend to behave similarly to local \textsc{Hii} regions, as highlighted in \cite{Strom_2018}

%Similar discrepancies when using N2O2 and N2S2 as tracers of N/O were highlighted by \cite{Strom_2018}, wherein high-redshift galaxies are seen to be offset from the local trend between the two diagnostics, instead behaving more similarly to local \textsc{Hii} regions. They proposed that the integrated light spectrum of local galaxies may include contributions from diffuse ionised gas (DIG) in addition to emission from \textsc{Hii} regions. Galaxies that have a larger contribution from DIG would be expected to have enhanced flux contribution to the [\ion{S}{II}] doublet, which would depress N2S2 at a fixed N2O2. If these galaxies are less dominated by \textsc{Hii} regions, which trace recent star formation, then we would also expect them to have a lower SFR, which suggests that the trend observed in Figure \ref{fig:N2O2vN2S2} may be driven by different contributions from DIG. However, recent work by \cite{Mannucci_2021} suggests that SDSS galaxies do not have a significant contribution from DIG, and instead differences in the measurements of low-ionisation line fluxes, such as the [\ion{S}{II}] doublet, are driven by aperture effects in local \textsc{Hii} regions.

%Due to the caveats discussed above, we caution the use of the N2S2 diagnostic to trace N/O, particularly for high-$z$ galaxies and \textsc{Hii} regions. Coupled with the relatively low detection rate of the full [\ion{S}{II}] doublet for our KLEVER sample, we prefer to adopt throughout this paper only the N2O2 diagnostic, which represents a far more direct tracer of N/O. 

\section{Results and Discussion}\label{Results}

\subsection{The relationship between O/H and N/O}

\subsubsection{SDSS ($z\sim$0)}
\begin{figure}
\includegraphics[width=\columnwidth]{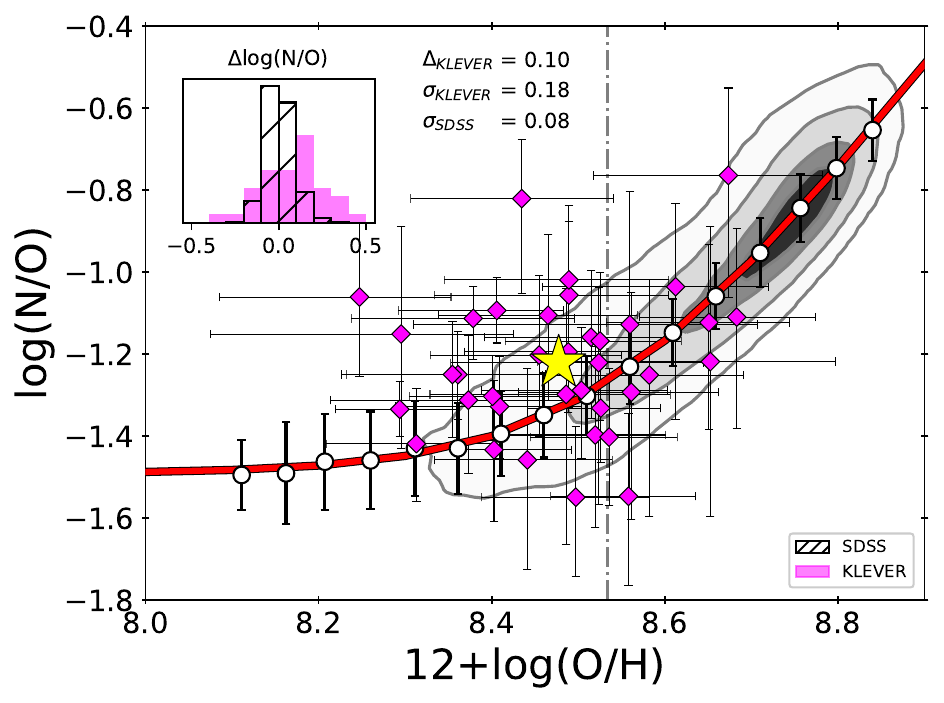}
\caption{The nitrogen abundance as a function of the gas-phase metallicity for the SDSS and KLEVER galaxy samples. The grey contours show the regions encompassing 30\%, 60\%, 90\% and 99\% of the local SDSS sample. The white points show the median binned SDSS data with corresponding standard deviations, with the solid red curve representing the best fit to the data using Eq. \ref{NOOH_eq}. The black dash-dotted line represents the transition metallicity at 12+log(O/H) = 8.534. The KLEVER galaxies are shown in magenta with corresponding errorbars, with the average position of the KLEVER galaxies shown as a yellow star. The histogram in the top left shows the deviations of the SDSS sample and the KLEVER sample from the best-fit line to the local data. KLEVER galaxies have a large scatter, yet on average show a mild enrichment in N/O of 0.1 dex when compared to SDSS galaxies.}
\label{fig:SDSS_KLEVER_NOOH}
\end{figure}

We begin by investigating the distribution of our SDSS galaxy sample in the 12 + log(O/H) vs log(N/O) plane, shown as grey contours in Figure \ref{fig:SDSS_KLEVER_NOOH}. The contours show the regions encompassing 30\%, 60\%, 90\% and 99\% of the total sample respectively. The white circles show the median nitrogen abundances taken in 0.05 dex bins of metallicity, with the errorbars representative of the standard deviations in N/O within each bin. The trend of N/O steeply rising at high metallicities and flattening at low metallicities has been observed commonly in several other studies in the past (\citealt{VilaCostas_1993}; \citealt{vanZee_1998}; \citealt{HenryWorthy_1999}; \citealt{PMC_2009}; \citealt{Pilyugin_2012}; \citealt{AndrewsMartini_2013}).
We choose the following functional form for parameterising the median N/O-O/H relation:
\begin{equation}\label{NOOH_eq}
\text{log(N/O)} = \text{log(N/O)}_0 + \gamma/\beta * \log\left(1 + \left(\frac{\text{(O/H)}}{\text{(O/H)}_0}\right)^\beta\right).
\end{equation}
In this equation, log(N/O)$_0$ represents the value of the nitrogen abundance plateau, where nitrogen production is considered primary and does not increase strongly with metallicity. At a metallicity of log(O/H) $>$ log(O/H)$_0$, the function is a simple power law of index $\gamma$, representing the region where secondary nitrogen production now dominates and N/O abundance increases strongly with metallicity. The sharpness of transition between the primary and secondary regimes is controlled by the $\beta$ parameter, with a smaller value of $\beta$ corresponding to a broader transition. A least-squares fit was performed to fit this model to the median binned data using the LMFIT package \citep[][]{Newville_2016}, with the fit being weighted by the standard error calculated using the dispersion in log(N/O) and number of objects within each bin. The best-fit parameter values and standard errors provided by LMFIT are given in Table \ref{tab:NOOH_best}. 

\begin{table}
 \centering
 \caption{The values of the best-fit parameters of the median N/O-O/H relation for the SDSS galaxy sample, fit using Eq. \ref{NOOH_eq}. Note that 12+log(O/H)$_0$ = 8.534 corresponds to a value of (O/H)$_0$ = 3.4 $\times$10$^{-4}$.}
 \label{tab:NOOH_best}
 \begin{tabular}{cccc}
  \hline
  \hline
  log(N/O)$_0$ & 12+log(O/H)$_0$ & $\gamma$ & $\beta$\\
  \hline
  $-$1.491 $\pm$ 0.017 & 8.534 $\pm$ 0.011 & 2.69 $\pm$ 0.11 & 3.6 $\pm$ 0.3\\[2pt]
  \hline
  \hline
 \end{tabular}
\end{table}

The best-fit to the SDSS data is shown as a solid red line in Figure \ref{fig:SDSS_KLEVER_NOOH}. We obtain a value for the low-metallicity plateau, of log(N/O)$_0$ $\approx$ $-$1.5, which is in good quantitative agreement with empirical estimates from other studies of \textsc{Hii} regions in nearby galaxies (\citealt{VilaCostas_1993}; \citealt{vanZee_1998}; \citealt{HenryWorthy_1999}; \citealt{PMC_2009}; \citealt{Pilyugin_2012}; \citealt{AndrewsMartini_2013}). The vertical dot-dashed line in Figure \ref{fig:SDSS_KLEVER_NOOH} shows the "transition" metallicity above which secondary nitrogen production dominates, found to be 12+log(O/H)$_0$ = 8.534$\pm$0.011, in good agreement with the value estimated in \cite{vanZee_1998} and \cite{AndrewsMartini_2013}. Other studies such as \cite{Shields_1991} and \cite{HenryWorthy_1999} suggest a transition metallicity lower by roughly 0.2 dex, at 12+log(O/H) = 8.3, whilst a polynomial fit to the \cite{Pilyugin_2012} presented in \cite{Steidel_2014} suggests a lower transition metallicity still at 12+log(O/H) = 8.2. One reason for the discrepancy in transition metallicities is the differences in parameterisation of the relationship between N/O and O/H. Some works have suggested a sharp transition between primary and secondary nitrogen production (e.g. \citealt{Charlot_2001}; \citealt{AndrewsMartini_2013}), whilst others present a more gradual transition between the two regimes (e.g. \citealt{vanZee_1998}; \citealt{Pilyugin_2012}). We show that a smooth transition is a more physical representation of the relationship, as the onset of significant secondary nitrogen production within different galaxies will vary dependent on their star formation and chemical enrichment histories. Nonetheless, the main cause of the varying estimations of transition metallicity values is the use of different calibrations and diagnostics to estimate the metallicity, as discussed in Section \ref{OH_det}.

Overall, we find that the relationship between N/O and O/H for our local sample has a scatter of just 0.08 dex for individual galaxies about the best-fit, comparable to that of the mass-metallicity relation (\citealt{Curti_2020FMR}). The overwhelming majority of our local galaxies (91\%) fall into the regime where secondary N production dominates (12+log(O/H)>8.538) and N/O is strongly correlated with O/H with a Spearman's rank correlation coefficient of 0.9. If we consider the remaining 9\% of the sample that have 12+log(O/H)<8.538, the correlation between N/O and O/H is significantly weakened, with the Spearman's rank coefficient of just 0.47 when only these galaxies are considered. This manifests in the increased scatter of the relationship at low metallicities, where the standard deviation of individual galaxies about the best-fit increases to 0.1 dex. 

\subsubsection{KLEVER ($z\sim$2.2)}

The distribution of our high-redshift sample in the N/O-O/H plane can be seen in Figure \ref{fig:SDSS_KLEVER_NOOH}. KLEVER galaxies are typically found at lower metallicities, with a median metallicity of 12+log(O/H) = 8.46 compared to the SDSS median of 12+log(O/H) = 8.72, which is expected from the evolution of the mass-metallicity relation at high-redshift and by the fact that the KLEVER galaxies are more highly star-forming and therefore, according to the FMR as discussed in Section \ref{Intro}, are expected to have lower metallicity. It is more interesting to investigate whether KLEVER galaxies at high redshift deviate from the N/O-O/H relation of local galaxies, as traced by the SDSS relation. The distribution of KLEVER galaxies appears more dispersed than the distribution of local galaxies. This is shown more clearly in the inset histogram in Figure \ref{fig:SDSS_KLEVER_NOOH} which shows the distribution of the deviations from the local best fit both for the SDSS galaxies and for the KLEVER galaxies. The dispersion of the KLEVER galaxies around the local relation is 0.18 dex, while the dispersion from SDSS galaxies is only 0.08 (although it increases to 0.10 at lower metallicities). 

%Taking the difference between these uncertainties, the additional dispersion in the KLEVER galaxies compared to SDSS is of the order 0.16 dex. 

The large scatter of high-redshift galaxies in the N/O-O/H plane has also been reported in \cite{Strom_2018}, where the distribution of high-$z$ galaxies was found to be well matched to the contours of local extra-galactic \textsc{Hii} regions from \cite{Pilyugin_2012}. If the large scatter of high-redshift galaxies in this plane is physically motivated, then it may indicate that the processes regulating the chemical enrichment of galaxies at high redshift are subject to a more varied range of phenomena that occur on different timescales (given the timescale sensitivity of N/O) with respect to local galaxies. We perform a simple test to examine how much of the dispersion observed in our KLEVER sample is due to the uncertainties in the measurement of metallicities and nitrogen abundances. We assume that the KLEVER galaxies fall on the best fit (red) line to the SDSS data, before perturbing both the nitrogen abundance and metallicity randomly within their measured errors. We then calculate the dispersion of the perturbed KLEVER sample relative to the local best fit, as done for the true KLEVER sample. We repeat this 1,000 times, finding that the mean dispersion of the perturbed KLEVER data is 0.2 dex. We do not include the intrinsic scatter in this test, but note that its inclusion would act to broaden the perturbed sample. The dispersion of 0.2 dex is comparable to the magnitude of the observed scatter, 0.18 dex, suggesting that the larger dispersion in KLEVER galaxies compared to the SDSS sample is expected due to the uncertainties in the estimation of N/O and O/H for our high-$z$ sample. 

In Figure~\ref{fig:SDSS_KLEVER_NOOH} we also show the average position of KLEVER galaxies in the N/O-O/H plane at log(N/O) = $-$1.22 and 12+log(O/H) = 8.46. Our high-redshift KLEVER galaxies tend towards larger N/O values at a fixed O/H compared to local galaxies, with an average offset of 0.1 dex in log(N/O). The standard error on the mean log(N/O) value is 0.03 dex, suggesting that the offset of 0.1 dex is significant. However, the large scatter of individual objects remains, and even with the slight offset of KLEVER galaxies to larger N/O values, the majority of the KLEVER sample still lie within the 99\% density contours of the SDSS sample, suggesting no strong evolution of the N/O-O/H relation. Our observed offset is smaller than that presented in \cite{Masters_2014} (M14), who report an average log(N/O) $= -$0.99 for a stacked sample of 26 high redshift galaxies. The difference in the absolute values of the nitrogen abundances between M14 and this paper are primarily driven by the choice of different calibrations for N/O. Indeed, if we adopt the calibration provided by \cite{Thurston_1996} which uses photoionisation codes to estimate the N/O ratio from strong emission lines, as in M14, our average N/O value increases to log(N/O) $\sim$ $-$1.1. M14 also report 0.4 dex enhancement in N/O at a fixed O/H, much larger than our own estimated enhancement. However, several factors make a direct comparison of the reported enhancements difficult, including the choice of different metallicity diagnostics and different local samples to compare to, e.g. whilst we use SDSS galaxies, M14 uses local \textsc{Hii} regions from \cite{vanZee_1998}. It is therefore more straightforward to compare between samples using line ratios, such as N2O2, directly. \cite{Shapley_2015} find that $z\sim$2 galaxies are offset to larger values of log(N2O2) by 0.24 dex when compared to SDSS galaxies near log(R3)$\sim$0.5. If we perform a similar analysis, we find that our KLEVER galaxies show a similar enhancement of 0.2 dex in log(N2O2). Similarly, \cite{Steidel_2016} present a composite spectrum of high-redshift galaxies from the Keck Baryonic Structure Survey (KBSS) with log(N2O2) = $-$1, which is 0.17 dex lower than the KLEVER average of log(N2O2) = $-$0.83. However,  the KBSS composite spectrum has log(O3N2) = 1.64, whereas for our KLEVER sample we find an average log(O3N2) = 1.06, where log(O3N2) $\equiv$ log(R3) - log(N2). The difference in sample selection between KBSS and KLEVER means that the difference in O3N2, and hence N2O2, is expected. Whilst KLEVER galaxies are H$\alpha$ selected, the KBSS sample differs in that it is rest-UV selected, meaning it may be biased against more massive galaxies, hence probing lower metallicities on average than our sample. The KBSS sample also requires slit-loss corrections in order to compare emission lines between different bands (such as H$\alpha$ and H$\beta$). These corrections have a median uncertainty of 8\%, introducing an additional source of error that is not present for the KLEVER sample. Our average values for N2O2 are in good agreement with the KBSS sample when compared in the log(O3N2) v log(N2O2) plane as presented in \cite{Strom_2017}. This suggests the level of nitrogen enrichment in our KLEVER sample is comparable to that of other high-redshift samples found in literature.

\subsection{N/O and stellar mass}

%\subsubsection{SDSS (z$\sim$0}

Nitrogen abundance is also known to correlate with the stellar mass of galaxies, with higher stellar masses having greater N/O ratios (\citealt{PMC_2009}). It has been argued by \cite{AndrewsMartini_2013} that this relation may be tighter than the N/O-O/H relation, however we note that their analysis relies on SDSS galaxies that have been stacked using stellar mass, a choice of binning which will artificially reduce the scatter in the N/O-M$_*$ plane. 

\begin{figure}
\includegraphics[width=\columnwidth]{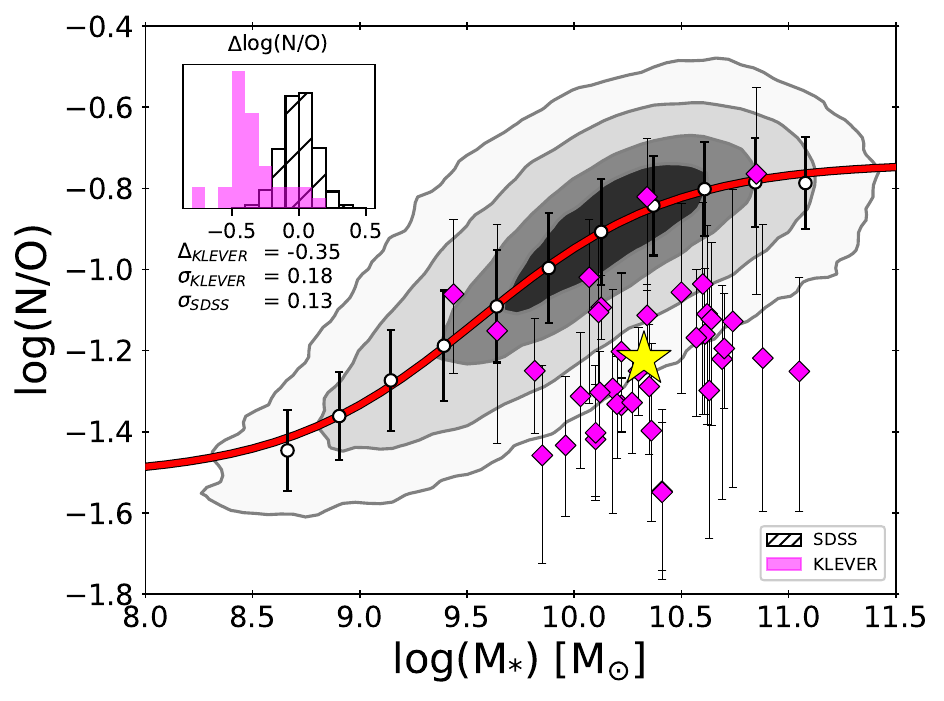}
\caption{The nitrogen abundance as a function of the stellar mass for the SDSS and KLEVER galaxy samples. The grey contours show the regions encompassing 30\%, 60\%, 90\% and 99\% of the total SDSS sample. The white points show the median binned SDSS data with corresponding standard deviations, with the solid red curve representing the best fit to the data using Equation \ref{NOmass_eq}. The magenta points are the KLEVER galaxies with corresponding errorbars, with the average position of the KLEVER galaxies shown as a yellow star. The histogram in the top left shows the deviations from the best fit in log(N/O) for both the KLEVER and SDSS samples. KLEVER galaxies have an average depletion in N/O of 0.35 dex compared to SDSS galaxies, suggesting significant redshift evolution in the stellar mass-N/O relation.}
\label{fig:SDSS_KLEVER_NOMass}
\end{figure}

An advantage of studying N/O against stellar mass rather than metallicity is that it provides a more "closed box" diagnostic for the chemical enrichment within a galaxy. This is because N/O and stellar mass are generally invariant to gas inflows and outflows, as opposed to the metallicity which depends more strongly on these processes. \cite{Masters_2016} argued that an enhanced N/O at fixed O/H could be a symptom of high-$z$ galaxies having higher stellar masses than local galaxies at a fixed O/H due to the redshift evolution of the mass-metallicity relation. If there is no strong evolution in the relationship between stellar mass and N/O, then this would cause high-$z$ galaxies to naturally have larger N/O values than their local counterparts.Mild evolution of the M$_*$-N/O relation has been found when using the N2S2 diagnostic to determine N/O, with high-$z$ galaxies tending towards lower N/O out to $z$ $\sim 0.4$ (\citealt{Perez-Montero_2013}) in a way expected if N/O tracks O/H. Similarly, \cite{Masters_2016} brought together data from \cite{Kashino_2017} and \cite{Steidel_2014} to show a mild evolution of N/O to values that are $\sim$0.1 dex lower at a fixed stellar mass out to $z$ $\sim$ 1.6. Yet, as already discussed in Section~\ref{NO_det}, determining the N/O abundance by using the N2S2 is affected by issues plaguing this diagnostic. However, works that instead use the N2O2 diagnostic to determine N/O such as \cite{Strom_2017} find a more significant evolution in the stellar mass-N/O relation, with N/O values up to 0.5 dex lower at a fixed stellar mass. 

\begin{figure*}
    \centering
    \begin{multicols}{2}
    \includegraphics[width=\columnwidth]{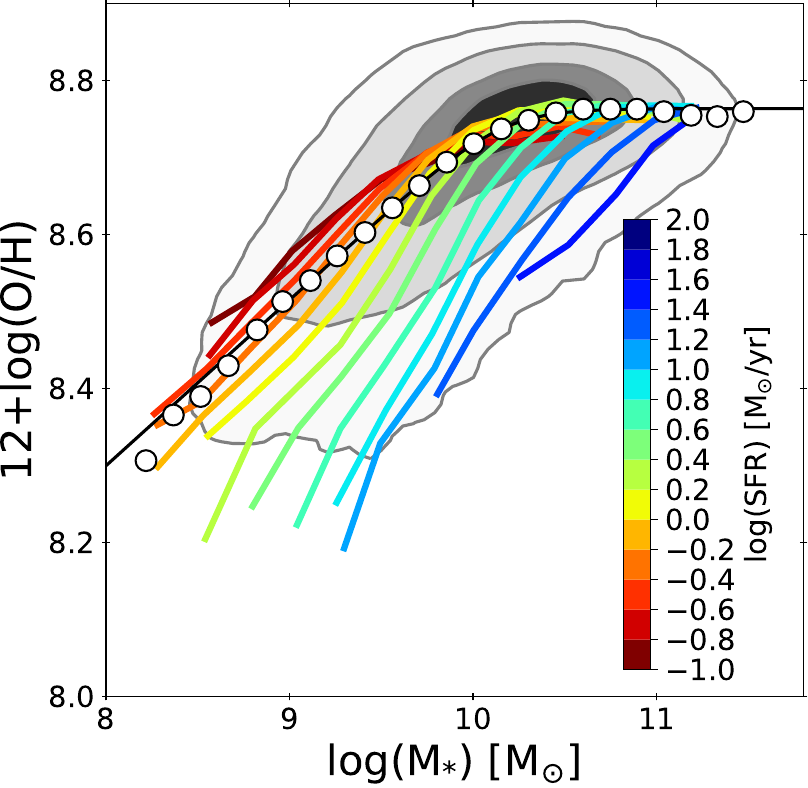} \par
    \includegraphics[width=\columnwidth]{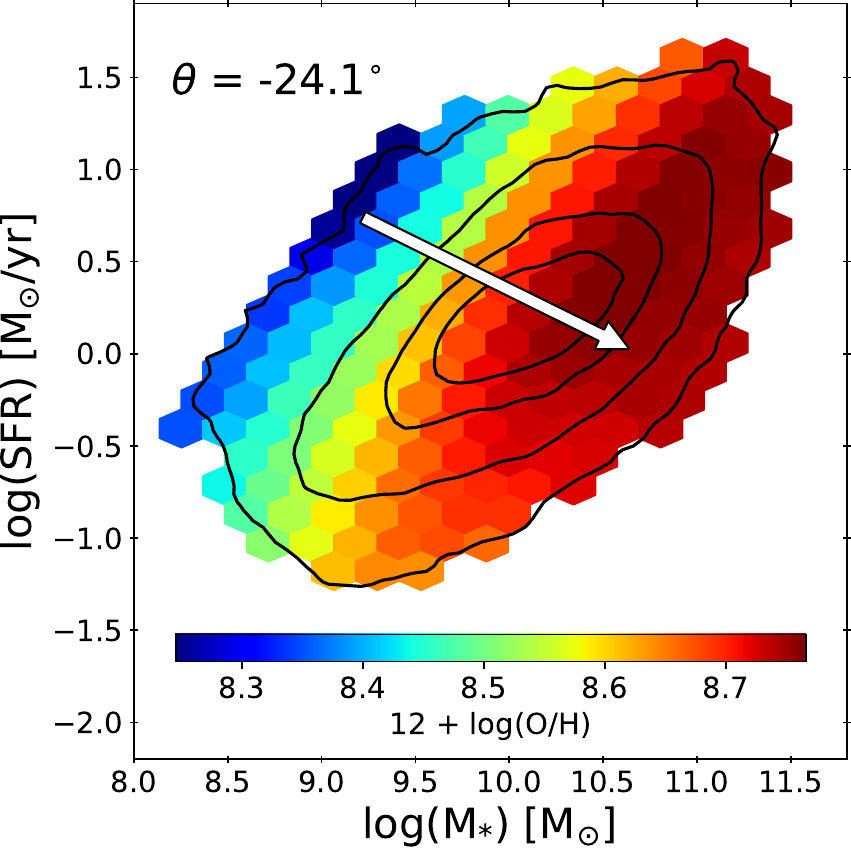} \par
    \end{multicols}
    \caption{Two representations of the SFR dependence of the MZR in local galaxies. The contours show the regions encompassing 30\%, 60\%, 90\% and 99\% of the sample. \textbf{Left}: The SDSS data has been binned into 0.2 dex wide bins of SFR. For each SFR bin, the median metallicity is measured in 0.25 dex wide bins of stellar mass, and all bins containing at least 25 objects are plotted. The median bins for the total sample are shown as white circles, with average MZR for the total sample is plotted as a black line. \textbf{Right}: the SDSS data has been spatially binned, with all bins containing at least 25 objects being shown, and the colour representing the average metallicity within each bin. The white arrow shows the direction in which metallicity is most efficiently increased for the total sample, derived using partial correlation coefficients as described in Section \ref{FMR_sec}. The angle of $-$24.1$^{\circ}$ highlights that the correlation between O/H and stellar mass is more than twice as strong as the anti-correlation between O/H and SFR.}
    \label{fig:all_FMR}
\end{figure*}

We have investigated these aspects in the SDSS and KLEVER samples in a homogeneous and consistent way, making use of our revised N/O calibration.
The trends between stellar mass and N/O for our SDSS and KLEVER samples are shown in Figure \ref{fig:SDSS_KLEVER_NOMass}. In order to parametrise a best fit to the local relation, we adopt an `S-shaped' function that accounts for a low log(N/O) plateau at low stellar masses, log(N/O)$_0$, as observed in \cite{AndrewsMartini_2013}, as well as a second log(N/O) plateau at high stellar masses, log(N/O)$_1$, to account for a flattening of the relation as observed in \cite{PMC_2009}. The function is then defined as
\begin{equation}\label{NOmass_eq}
\text{log(N/O)} = \text{log(N/O)}_1 - \frac{\left[\text{log(N/O)}_1 - \text{log(N/O)}_0\right]}{\left[ 1 + \left( M_* / M_0\right) ^k \right]},
\end{equation}
where $k$ defines the steepness of the transition between the low and high N/O plateaus and M$_0$ is the stellar mass about which the gradient of the function begins to decrease. Similar to our method for the N/O-O/H relation, we fit this function by binning in 0.25 dex wide intervals of stellar mass, taking the median nitrogen abundance and calculating the standard deviation within each bin (shown as white circles in Figure \ref{fig:SDSS_KLEVER_NOMass}). We then performed a least-squared regression to fit Equation~\ref{NOmass_eq} to the median-binned data, including only bins containing at least 25 galaxies and weighting by the standard error calculated from the standard deviation in N/O and the number of objects within each bin. The best fit parameters are reported in Table \ref{tab:NOMass_params}. 
\begin{table}
 \centering
 \caption{The values of the best-fit parameters of the median N/O-stellar mass relation for the SDSS galaxy sample, fit using Eq. \ref{NOmass_eq}.}
 \label{tab:NOMass_params}
 \begin{tabular}{cccc}
  \hline
  \hline
  log(N/O)$_0$ & log(N/O)$_1$ & $k$ & log(M$_0$/M$_\odot$)\\
  \hline
  $-$1.51 $\pm$ 0.04 & $-$0.737 $\pm$ 0.015 & 0.95$\pm$ 0.08 & 9.55 $\pm$ 0.05\\[2pt]
  \hline
  \hline
 \end{tabular}
\end{table}
The resulting fit is shown as a solid red curve in Figure \ref{fig:SDSS_KLEVER_NOMass}, and appears visually to be a good parametrisation of our binned SDSS data. Whilst we do not probe many galaxies below a stellar mass of log(M/M$_\odot$) = 8.9, the proposed end of the low N/O plateau found in \cite{AndrewsMartini_2013}, we find that the value of the plateau is in good agreement with the value found earlier in the N/O-O/H plane at log(N/O) $\sim$ $-$1.5. At the high mass end the relationship is found to flatten at a value of log(N/O)$\sim$ $-$0.74, which could be interpreted in the context of the flattening of the mass-metallicity relation at high stellar masses (\citealt{Tremonti_2004}). When O/H is no longer increasing with stellar mass, the strong dependency of N/O on O/H means a flattening in O/H with stellar mass will be reflected in N/O. For our SDSS galaxy sample the relationship between N/O and stellar mass is less tight than the N/O-O/H relation, with a scatter of 0.13 dex in N/O about the median relation for individual galaxies. 

%\subsubsection{KLEVER (z$\sim$2.2)}

The KLEVER galaxies are shown in the stellar mass-N/O plane in Figure \ref{fig:SDSS_KLEVER_NOMass}, along with the SDSS comparison sample. The positive correlation between stellar mass and N/O is still recovered at $z$ $\sim$ 2, with a Spearman's rank correlation coefficient of 0.31 and p = 0.06, although the correlation is less strong than that found in the SDSS sample, where the Spearman's rank correlation coefficient is 0.74. Our high-redshift galaxies clearly occupy a different region of the N/O-stellar mass plane than the local SDSS galaxies, with roughly half of the KLEVER sample falling below the 99\% density contours of the local sample. We find that, on average, KLEVER galaxies fall 0.35 dex below the local best-fit relation. This suggests a significant evolution in the relationship between stellar mass and N/O between $z$ $\sim$ 0 and $z$ $\sim$ 2. Our results are in agreement with those from \cite{Strom_2017}, who find that KBSS galaxies at $z$ $\sim$ 2.3 are $\sim$0.32 dex lower in N/O at a fixed stellar mass. The strong evolution of the stellar mass-N/O relation is comparable to the evolution of the mass-metallicity relation, where studies have observed a change in O/H of $-$0.26 dex at a fixed stellar mass for galaxies at $z$ $\sim$ 2 compared to local samples (\citealt{Sanders_2020FMR}). Such similar evolution of the mass-metallicity relation and the stellar mass-N/O relation could suggest that the N/O-O/H plane is approximately redshift-invariant.

In contrast with our result, 
\cite{Perez-Montero_2013} and \cite{Masters_2016} found the mass-N/O relation to be redshift invariant or subject to a more modest redshift evolution.
Beyond the different redshifts probed, $z\sim$0.4 for \cite{Perez-Montero_2013} and $z\sim$1.6 in \cite{Masters_2016}, the main reason for the difference with our result is their choice of N2S2 as a diagnostic, rather than N2O2. As highlighted in Figure 12 of \cite{Strom_2018}, there is an offset in the relationship between N2O2 and N2S2 between local galaxies and high-redshift galaxies, which behave more like \textsc{Hii} regions. This offset will mean that using N2S2 to directly trace the nitrogen abundance without using an appropriate conversion to N/O, as done in \cite{Masters_2016} and \cite{Kashino_2017}, will lead to N/O being underestimated in local galaxies with respect to high-redshift samples, making any evolution in the stellar mass-N/O plane less apparent. The use of N2S2 as a diagnostic for the nitrogen abundance and the physical interpretation of evolution in the stellar mass-N/O plane are discussed further in Section \ref{FNR_sec}.

\subsection{The Fundamental Metallicity Relation (FMR)}

\subsubsection{SDSS ($z\sim$0)}\label{FMR_sec}
Several works have investigated the secondary dependence of the mass-metallicity relation on star formation rate, finding that the relationship between these three variables may also be redshift invariant out to $z$ $\sim$ 3 (\citealt{Mannucci_2010}; \citealt{AndrewsMartini_2013}; \citealt{Cresci_2019}; \citealt{Curti_2020FMR}; \citealt{Sanders_2020FMR}). Two 2D representations of the local relationship between M$_*$-O/H-SFR are shown in Figure \ref{fig:all_FMR}. In the left-hand panel, the contours and shaded regions show the density of SDSS galaxies in the mass-metallicity plane, while coloured lines show the trend of SDSS galaxies in 0.2 dex wide bins of SFR. For each bin, we take the median O/H in 0.25 dex wide intervals of stellar mass, plotting all bins containing at least 25 objects. The average MZR for the full SDSS sample is shown as a black line, with points of median O/H across the range of stellar masses probed shown as white circles. The majority of the scatter below the average mass-metallicity relation can be explained by variations in star formation rate, with the individual SFR tracks extending out to the 99\% density contours of the SDSS sample. The correlation between M$_*$-O/H-SFR is often understood in the context of gas inflows, where infalling pristine gas dilutes the metallicity whilst fostering star formation at a specific stellar mass (\citealt{Dave_2011}). In this scenario, the secondary anti-correlation between metallicity and SFR should be a byproduct of a more fundamental relation between metallicity and gas content (\citealt{Bothwell_2013}, \citeyear{Bothwell_2016a}, \citeyear{Bothwell_2016b}), although with lower statistics.  Other factors may also play a role in causing the fundamental metallicity relation, such as metal-loaded outflows, gas recycling through galactic fountains, or variations in the IMF (\citealt{Tremonti_2004}; \citealt{Koppen_2007};\citealt{Dave_2011}; \citealt{Sanders_2020FMR}).

\begin{figure*}
    \centering
    \includegraphics[width=\linewidth]{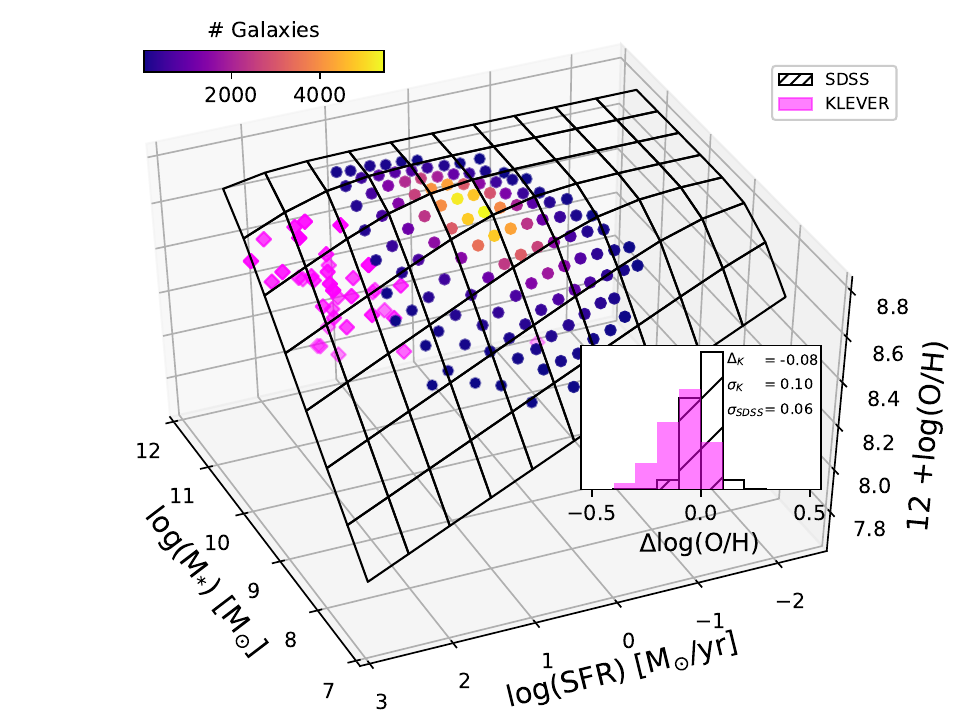} \par
    \caption{The 3D FMR showing the relationship between M$_*$-O/H-SFR. The grid shows the relationship derived from the binned SDSS sample, shown as circles, with the colourbar showing the number of galaxies within each M$_*$-SFR bin.} KLEVER galaxies are shown in magenta, and the difference in the scatter of individual galaxies about the FMR for both SDSS and KLEVER galaxies is shown in the histogram. The average offset of KLEVER galaxies from the FMR is just 0.08 dex, which suggests that KLEVER galaxies lie on the FMR within uncertainties.
    \label{fig:FMR_3D}
\end{figure*}

In Figure \ref{fig:all_FMR}, we also show an alternative but equivalent representation of the relationship between M$_*$-O/H-SFR. Here, we spatially bin the SDSS sample in the mass-SFR plane, displaying all bins that contain at least 25 objects, with the colour representing the average metallicity within each bin. The white arrow shows the direction of the metallicity gradient in the stellar mass-SFR plane, as calculated using partial correlation coefficients (PCCs). Partial correlation coefficients quantify the strength of correlation between two variables, A and B, when keeping a third variable, C, fixed. This is defined by
\begin{equation}\label{PCC_er}
\rho_{AB,C} = \frac{\rho_{AB} - \rho_{AC}\cdot\rho_{BC}}{\sqrt{1-\rho^{2}_{AC}}\sqrt{1-\rho^{2}_{BC}}},
\end{equation}
where $\rho_{AB}$ represents the Spearman's rank correlation coefficient between variables A and B. Following \cite{Bluck_2020}, we can then take the PCC between O/H and M$_*$ fixing SFR and the PCC between O/H and SFR fixing M$_*$, and treat them as two components of a vector in order to determine the direction in which we must move on the mass-SFR plane to maximise the change in O/H. This direction is given by
\begin{equation}\label{PCC_vector_eq}
\theta = \arctan\left(\frac{\rho_{Yz, X}}{\rho_{Xz, Y}}\right),
\end{equation}
where $\theta$ is measured from the horizontal, $X$ and $Y$ are the variables on the respective $x$ and $y$ axes, and $z$ is the third variable being investigated. If the anti-correlation between O/H and star formation rate is equally as strong as the correlation between O/H and stellar mass, we would find that $\theta$ = $-$45$^{\circ}$. Instead, we determine that the angle that maximises the change in O/H on the stellar mass-SFR plane is $\theta$ = $-$24.1$^{\circ}$. This result highlights that the correlation between O/H and stellar mass is stronger than the anti-correlation between O/H and SFR, which has a secondary but significant role in regulating the O/H. We note that the fundamental metallicity relation found by \cite{Mannucci_2010}, in which the SFR accounts for about 1/3 of the metallicity variation would have implied an angle of -30$^{\circ}$ for the gradient arrow. Our results suggest a similar dependency, although the finding that the gradient arrow is flatter indicates that the SFR has a somewhat lesser role in regulating the metallicity. The difference between the result expected from \cite{Mannucci_2010} and our own may be driven in part by the use of different metallicity diagnostics.

%In the context of the FMR, this could be interpreted as the stellar mass being the dominant factor in determining a galaxies location on the FMR plane, with deviations from this position  caused by fluctuations in O/H and SFR linked to the inflow and outflow of gas as discussed earlier in this section.

\subsubsection{KLEVER ($z\sim$2.2)}

As larger samples of high-redshift galaxies have been assembled, several authors have studied whether the local fundamental metallicity relation still holds at high-redshift. Some works have shown that high-redshift galaxies often fall below the FMR relative to their expected position for a given stellar mass and SFR (\citealt{Cullen_2014}; \citealt{Troncoso_2014}; \citealt{Sanders_2015}, \citeyear{Sanders_2018}), although this offset is often small and of the order $\sim$ 0.1 dex. Such an offset is well within the predictive errors of locally calibrated FMR planes, especially when considering the potential evolution of metallicity diagnostics at high-$z$ (discussed in Section \ref{OH_det}). Indeed, recent work by \cite{Sanders_2020FMR} found that using a metallicity calibration from \cite{Bian_2018} for objects at $z>$2 yields metallicity estimates that are 0.05-0.1 dex larger than those made using locally calibrated diagnostics, effectively cancelling any offset.

We investigate whether our KLEVER galaxies fall on the fundamental metallicity relation using the parameterisation provided in \cite{Curti_2020FMR}, adopting total-SFR estimates for SDSS galaxies. The local FMR is shown as a 3D plane in Figure \ref{fig:FMR_3D}, with the locations of the KLEVER points shown in magenta. The inset in the same figure shows the distribution of galaxies around the FMR, both for SDSS local galaxies and for the KLEVER galaxies at $z\sim$2.2 . As can be seen, the average deviation of the O/H for KLEVER galaxies from that estimated using the stellar mass and SFR is just 0.08 dex. This offset is possibly due to our use of locally calibrated metallicity diagnostics, but nevertheless falls well within the uncertainties associated to metallicity predictions made using the FMR that are of the order $\sim$0.15 dex, as shown in Figure 11 of \cite{Curti_2020FMR}. KLEVER galaxies also probe a region of the FMR that is not populated by local galaxies, therefore any prediction of their metallicity using the parametrised form of the FMR relies on its extrapolation. As such, we conclude that, to the level of accuracy granted by the present observations, KLEVER galaxies are consistent with evolutionary scenario depicted by the FMR, suggesting that the processes governing the baryonic growth of galaxies operate in the same manner at $z$ $\sim$ 2 as they do locally.

\begin{figure*}
    \centering
    \begin{multicols}{2}
    \includegraphics[width=\columnwidth]{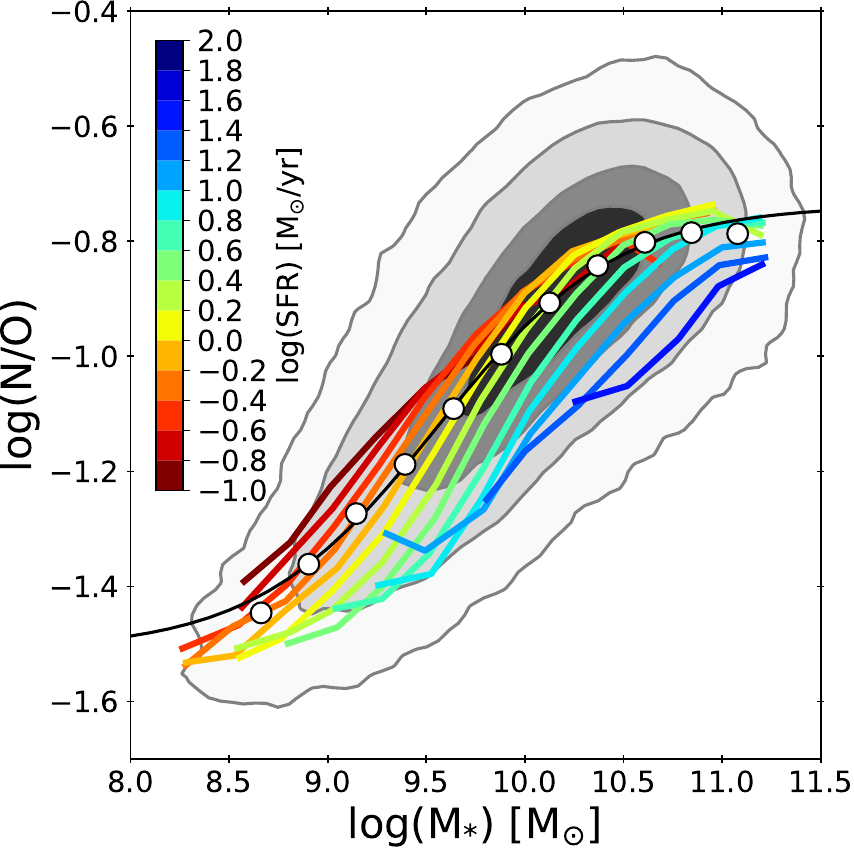} \par
    \includegraphics[width=\columnwidth]{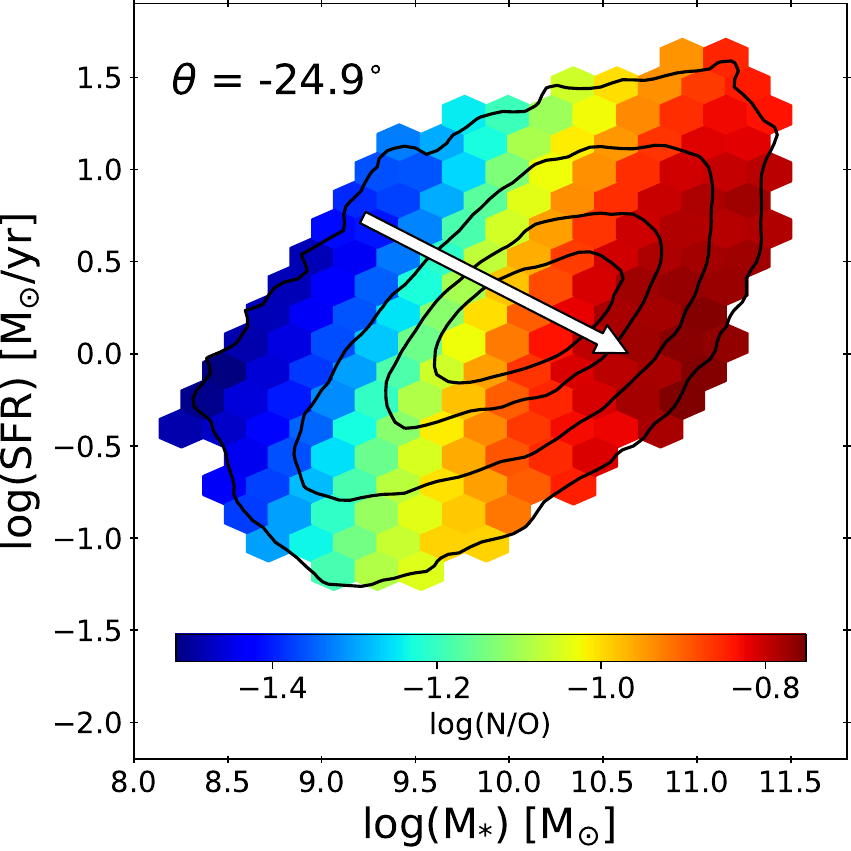} \par
    \end{multicols}
    \caption{Two representations of the SFR dependence in the stellar mass-N/O plane for local galaxies. The contours show the regions encompassing 30\%, 60\%, 90\% and 99\% of the sample. \textbf{Left}: The SDSS data has been binned into 0.2 dex wide bins of SFR. For each SFR bin, the median N/O is measured in 0.25 dex wide bins of stellar mass, and all bins containing at least 25 objects are plotted. The median bins for the total sample are shown as white circles, with average stellar mass-N/O relation for the total sample is plotted as a black line. \textbf{Right}: the SDSS data has been spatially binned, with all bins containing at least 25 objects being shown, and the colour representing the average nitrogen abundance within each bin. The white arrow shows the direction in which nitrogen abundance is most efficiently increased for the total sample, derived using partial correlation coefficients as described in Section \ref{FMR_sec}. The angle of $-$24.9$^{\circ}$ is extremely similar to the angle of $-$24.1$^{\circ}$ derived for the fundamental metallicity relation, highlighting the similarity of the M$_*$-SFR-O/H and M$_*$-SFR-N/O relations.}
    \label{fig:all_FNR}
\end{figure*}

\subsection{A Fundamental Nitrogen Relation (FNR)?}

\subsubsection{SDSS ($z\sim$0)}\label{FNR_sec}

Analogous to the FMR defining the relationship between M$_*$-O/H-SFR, we can replace O/H with N/O to probe the potential existence of a Fundamental Nitrogen Relation (FNR). N/O is not sensitive to many of the physical processes that O/H is, such as pristine gas inflows\footnote{We do note that N/O may be sensitive to inflows that are rich in oxygen, but do not contain much nitrogen. Such a scenario is compatible with the re-accretion of enriched gas that was expelled from a galaxy early in its evolution, before N enrichment could begin, and would have the effect of decreasing the N/O ratio as long as the inflowing gas mass is large compared to the gas mass within the galaxy. This dilution scenario will be explored in more detail in future work.}, and instead is sensitive to the enrichment timescales within the galaxy due to the delayed production of nitrogen relative to oxygen. This makes the relationship between M$_*$-N/O-SFR a powerful probe of galaxy evolution in regards to quantities relating to enrichment timescales, such as galaxy age and star formation efficiency.

We begin by investigating the relationship between M$_*$-N/O-SFR in the 2D stellar mass-N/O plane for local galaxies, following the analysis presented in Section \ref{FMR_sec}. The results can be seen in Figure \ref{fig:all_FNR} which is analogous to Fig.~\ref{fig:all_FMR}, where now O/H is replaced with N/O. The N/O ratio depends not only on stellar mass, but also on SFR (at fixed M$_{\star}$): indeed, an anti-correlation exists between N/O and SFR, at fixed stellar mass, similar to the anti-correlation between O/H and SFR found in the fundamental metallicity relation. The similarities between the anti-correlations of N/O and O/H with SFR are well highlighted by considering the partial correlation coefficient vectors on the right-hand side of Figures \ref{fig:all_FMR} and \ref{fig:all_FNR}. The direction to maximise the N/O variation in the stellar mass-SFR plane is $\theta$ = $-$24.9$^{\circ}$ below the horizontal, which is fully consistent with the value of $\theta$ = $-$24.1$^{\circ}$ found in the FMR.

We note that if the secondary dependence of O/H on SFR was due to accretion of pristine gas (diluting O/H and fostering SFR), then this should not translate into a secondary dependence of N/O on SFR, as N/O is not sensitive to accretion of pristine (H) gas. The finding that N/O has essentially the same dependence on SFR as O/H suggests that the accretion of pristine gas is not the only origin of the O/H-SFR anti-correlation. An alternative explanation could be that the accreting gas is metal poor, although not pristine, and slightly enriched by primary nitrogen (i.e. log(N/O) $\sim -1.5$). This scenario will be investigated in more detail in future work. Age and star formation history effects, which are linked to the current level of SFR, are also likely to play an additional important role in explaining both relations.

However, we also note that there are differences in the amount of scatter that can be explained by the SFR in these two planes. Unlike in the mass-metallicity relation, the star formation rate tracks in the stellar mass-N/O plane do not extend below the 90\% density contours of the local sample. Furthermore, the scatter above the average stellar mass-N/O relation is much larger than that above the MZR, meaning a fundamental nitrogen relation constructed form the SFR tracks will not improve the scatter in N/O as dramatically as the fundamental metallicity relation is able to for O/H. 

The relationship between M$_*$-N/O-SFR has previously been investigated by \cite{Perez-Montero_2013}, who found no strong dependence on SFR in the stellar mass-N/O plane, suggesting that the relationship between O/H and SFR is driven primarily by inflows of metal poor-gas, which N/O is insensitive to. Once again, our results are different mainly due to \cite{Perez-Montero_2013} using the N2S2 diagnostic to derive N/O, whilst we use N2O2, with the differences between this diagnostics highlighted earlier in Section~\ref{NO_det}. Indeed, if we adopt the N2S2 diagnostic for the sample in this paper and re-plot the correlation between stellar mass and N/O, we do not identify any dependence on SFR in the stellar mass-N/O plane, aligning our results with those of \cite{Perez-Montero_2013}. We note that the dependence on SFR, when using N2O2 as a proper diagnostic, is not unique to our own calibrations, and remains if the calibrations from PMC09 are used instead. This further highlights how the choice of diagnostic for N/O can impact the results.

Given the dependence on SFR that we observe in the stellar mass-N/O plane when adopting the N2O2 diagnostic, we suggest that the fundamental metallicity relation is not purely driven by the accretion of pristine gas, but must also be partially driven by varied chemical enrichment timescales.

\begin{figure*}
    \centering
    \includegraphics[width=\linewidth]{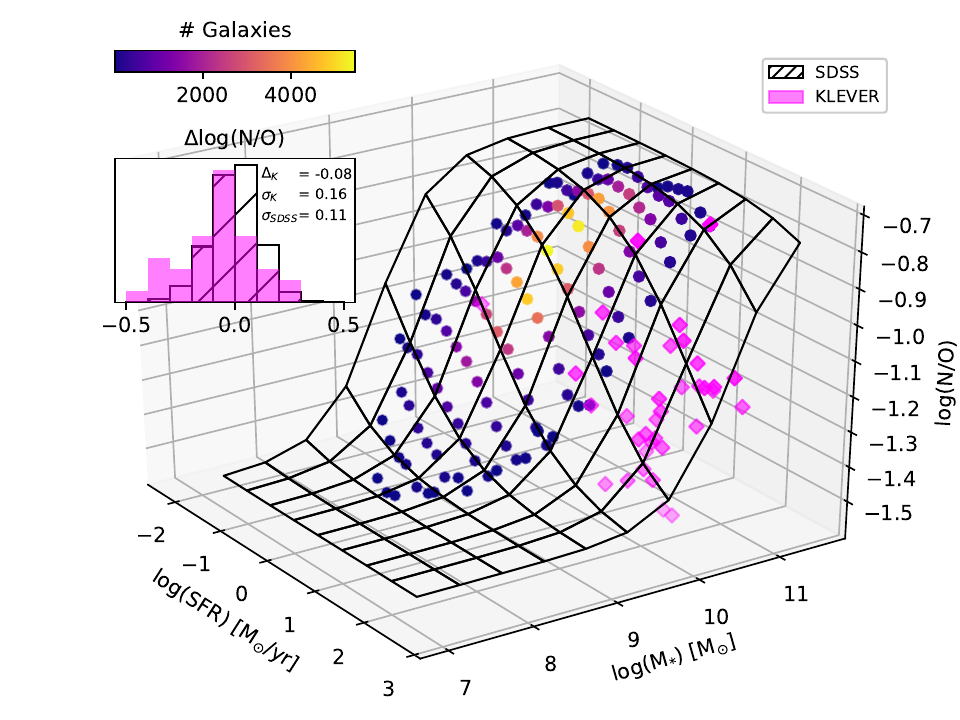} \par
    \caption{The 3D FNR showing the relationship between M$_*$-N/O-SFR as derived in this paper. The grid shows the relationship derived from the binned SDSS sample, shown as circles, with the colourbar showing the number of galaxies within each M$_*$-SFR bin.}. KLEVER galaxies are shown in magenta, and the difference in the scatter of individual galaxies about the FNR for both SDSS and KLEVER galaxies is shown in the histogram. Similar to the results presented in Fig~\ref{fig:FMR_3D}, the average offset of KLEVER galaxies from the FNR is just 0.08 dex, suggesting that KLEVER galaxies lie on the local fundamental nitrogen relation plane.
    \label{fig:FNR_3D}
\end{figure*}

To test whether the relationship that we observe between M$_*$-N/O-SFR evolves with redshift, we aim to define a relationship in the 3D plane between these variables, following a similar approach as for the fundamental metallicity relation in \cite{Curti_2020FMR}. We fit the SFR tracks from the left-hand panel of Figure \ref{fig:all_FNR} with the functional form of the stellar mass-N/O relation from Equation \ref{NOmass_eq} using a least-square minimisation. We fit only those SFR tracks with at least 8 data points, each of which contain at least 25 galaxies, to ensure the tracks probe a wide range of stellar masses. However, even with this requirement, not all SFR tracks extend to both the low and high mass regimes where we expect to see plateaus in N/O. As such we set the log(N/O)$_0$ value to the global value of $-$1.5 and the log(N/O)$_1$ value to $-$0.74 for all SFR tracks. The remaining two parameters, M$_0$ and $k$, are allowed to vary freely. There is a linear positive correlation between M$_0$ and SFR, whilst $k$ shows only a mild non-linear dependence on star formation rate, varying between values of 1 and 1.4. The higher turnover mass (M$_0$) for more star forming galaxies can be interpreted in the context of age of the galaxy and the star formation efficiency. If we consider SFR to be a tracer of galaxy age then at a fixed stellar mass younger, more star-forming galaxies have had less time for the delayed enrichment of nitrogen relative to oxygen to occur and so have lower N/O ratios. Similarly if we instead consider galaxies at a fixed N/O, the more star forming galaxies tend towards larger stellar masses. In order for these young galaxies to assemble a large stellar mass in short time, they must be efficient at converting their gas into stars, and thus must have larger star formation efficiencies. Therefore the observed trend could be due to a variation in galaxy age and star formation efficiency.

We obtain the final functional form for our 3D fundamental nitrogen relation by taking Equation \ref{NOmass_eq} and allowing M$_0$ to vary with SFR as log(M$_0$/M$_\odot$) = $m_1$ + $m_0$log(SFR), with $k$ allowed to vary freely. We then use a least-square minimisation to fit this function to the median binned SFR tracks in the stellar mass-N/O plane, with the final parameter values reported in Table \ref{tab:FNR_params}.
\begin{table}
 \centering
 \caption{The values of the best-fit parameters quantifying the variation of turnover mass with SFR and steepness of the transition from low to high N/O plateaus in our parameterisation of the FNR.}
 \label{tab:FNR_params}
 \begin{tabular}{cccc}
  \hline
  \hline
  $m_1$ & $m_0$ & $k$\\
  \hline
   9.584$\pm$ 0.004 & 0.451 $\pm$ 0.010 & 1.324 $\pm$ 0.019 \\[2pt]
  \hline
  \hline
 \end{tabular}
\end{table}
The resulting 3D representation of the fundamental nitrogen relation can be seen in Figure \ref{fig:FNR_3D}. The SDSS points are taken in bins of M$_*$ and SFR, and are shown as circles with the colour representing the number of galaxies in each bin.

\begin{figure*}
    \centering
    \begin{multicols}{2}
    \includegraphics[width=0.946\columnwidth]{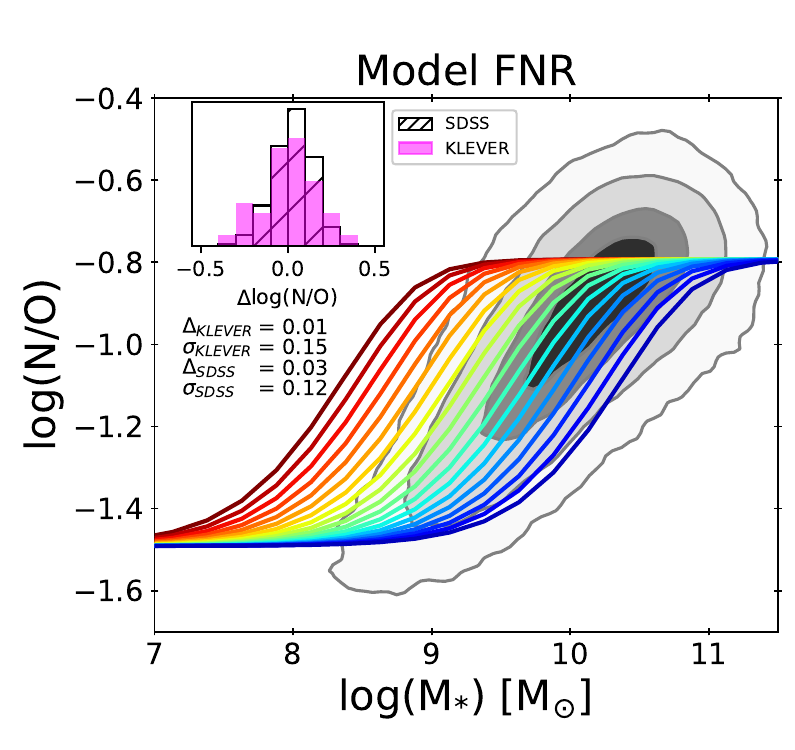} \par
    \includegraphics[width=\columnwidth]{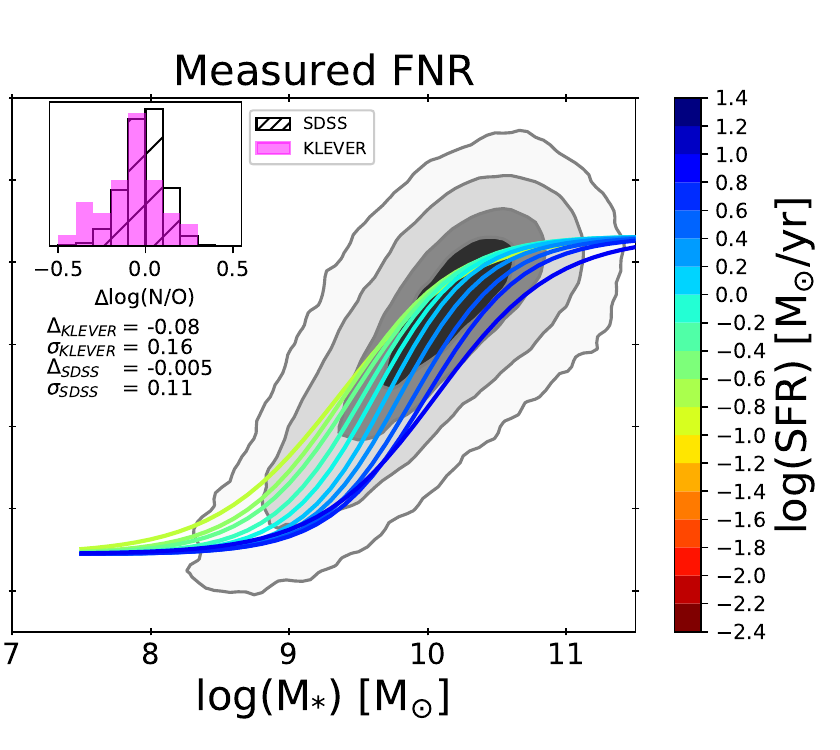} \par
    \end{multicols}
    \caption{The 2D projection of the fundamental nitrogen relation for both our model (left) and measured (right) relationships. The contours show the regions encompassing 30\%, 60\%, 90\% and 99\% of the SDSS sample. The coloured lines show tracks of constant SFR. For both the model and measured FNR, the inset histogram in the top left of each plot shows the difference between the expected and true log(N/O) values for both SDSS and KLEVER galaxies. The median offsets and standard deviations of the SDSS and KLEVER galaxies from both the model and measured FNR are shown below the inset histograms. The similar performance of the model and measured fundamental nitrogen relations suggests that the relationship between M$_*$-N/O-SFR is driven by a combination of the fundamental metallicity relation and a redshift-invariant N/O-O/H relation.}
    \label{model_FNR}
\end{figure*}

\subsubsection{KLEVER ($z\sim$2.2)}\label{FNR_klever}

 The location of KLEVER galaxies on the fundamental nitrogen relation 3D plane is shown in Figure \ref{fig:FNR_3D}. The inset histogram shows the distribution of the deviations of SDSS and KLEVER galaxies from the FNR plane. As can be seen, KLEVER galaxies are generally found to be scattered about the fundamental nitrogen relation, meaning that although nitrogen appears to be under-produced in high-$z$ galaxies relative to local ones when compared at a given stellar mass, if we account for the increase in star formation rate then KLEVER galaxies are well described by the local relationship between M$_*$-N/O-SFR. If we interpret the variation between M$_*$-N/O-SFR as being driven by time-related quantities such as galaxy age and star formation efficiency, then the high star formation rate values of the KLEVER galaxies, coupled with high stellar masses and low nitrogen abundances, suggests that KLEVER galaxies are chemically young and potentially have higher star formation efficiency than local galaxies (which results in faster oxygen enrichment relative to nitrogen). Furthermore, the apparent redshift invariance of the fundamental nitrogen relation suggests that any variations in M$_*$, N/O or SFR driven by changes in the age or star formation efficiency occur smoothly between $z$ $\sim$ 2 and $z$ $\sim$ 0. This interpretation is in agreement with the suggestion from \cite{Masters_2016} that a relation between N/O, M$_*$ and galaxy age may exist, with the dependence on star formation rate acting as a tracer of the evolutionary stage of galaxies in our fundamental nitrogen relation.

\subsubsection{The FNR--FMR symbiosis}

The existence of a fundamental relationship between M$_*$, N/O and SFR may be expected given the existence of both the fundamental metallicity relation and the N/O-O/H relation, assuming that these relationships hold for all galaxies. To compare our measured fundamental nitrogen relation to that which we would expect from a combination of the fundamental metallicity relation and the N/O-O/H relationship, we create a `model' FNR. Within our model FNR, for a given stellar mass and star formation rate we calculate the O/H expected from the paramaterisation of the fundamental metallicity relation presented in \cite{Curti_2020FMR}, and shown in Figure \ref{fig:FMR_3D}. We then convert this expected O/H value into a value of N/O using our median N/O-O/H relation, given by Equation \ref{NOOH_eq}. The results are shown in Figure \ref{model_FNR}, where in the left hand panel the expected N/O values from the model are plotted against stellar mass in tracks of constant SFR, whilst the right hand panel shows the same plot but with expected N/O values calculated from our measured fundamental nitrogen relation. The inset histogram in the top left of each panel shows the offset between the expected and measured values of log(N/O) for both the SDSS and KLEVER samples. 

The similarities between the model and measured FNRs is immediately visible, with the `S-shaped' form of the star formation rate tracks in the model FNR being naturally reproduced by a combination of the functional forms of the mass-metallicity relation and our N/O-O/H relationship, justifying our choice of Equation \ref{NOmass_eq} to parameterise the relationship between N/O and stellar mass. Both the model and measured FNR perform well in accurately predicting the N/O values of both the SDSS and KLEVER samples, with the offset in log(N/O) less than 0.1 dex in all cases. The precision of the model and measured FNRs are also comparable, with the scatters in log(N/O) = 0.12 and 0.11 for the SDSS sample and log(N/O) = 0.15 and 0.16 for the KLEVER sample respectively. The equivalency of our model and measured fundamental nitrogen relations suggest that the fundamental nitrogen relation is primarily driven by the combination of the fundamental metallicity relation with a redshift-invariant N/O-O/H relation. This suggests that evolution of the mass-metallicity relation is driven by processes that are probed not only by O/H, but also by N/O.  

\subsubsection{The role of galaxy evolutionary stage}

\begin{figure}
\includegraphics[width=\columnwidth]{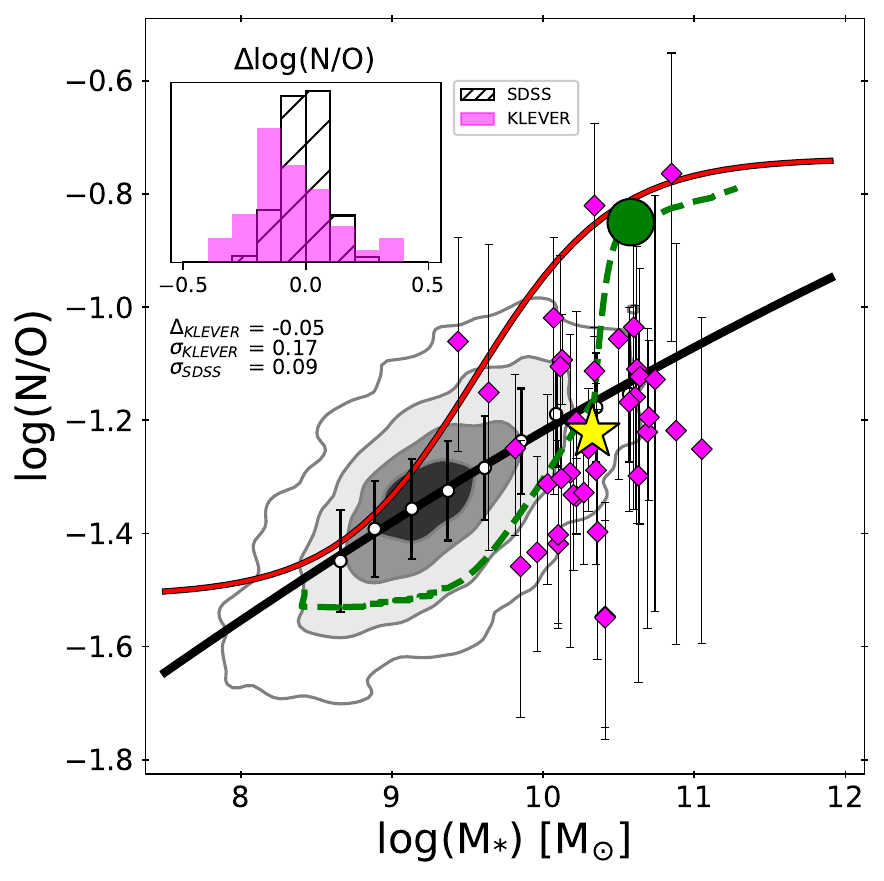}
\caption{N/O as a function of the stellar mass. The grey contours show the regions encompassing the 30\%, 60\% 90\% contours of the local SDSS galaxies with a gas phase metallicity of 12+log(O/H)<8.534. The magenta points represent individual KLEVER  galaxies, with the average position of the KLEVER galaxies shown as a yellow star. The solid red line shows the median N/O-M$_{*}$ relation derived for the full SDSS sample, whilst the solid black line shows the best fit to the median N/O-M$_{*}$ relation for only SDSS galaxies with 12+log(O/H)$<$8.534. The histogram in the top left shows the deviation in N/O from the black line for individual low metallicity SDSS galaxies and KLEVER galaxies. The dashed green line shows the evolutionary track for a single galaxy using chemical evolution models with an infall mass log(M$_{inf}$) = 12 M$_{\sun}$, infall time $\tau_{inf}$ = 7 Gyr, mass loading factor $\omega$ = 1.0 and star formation efficiency $\nu$ = 5 Gyr$^{-1}$ (\citealt{Vincenzo_2016}). The green circle shows the position of the galaxy at an age of 1 Gyr. The KLEVER galaxies are offset from the local relation by only $-$0.05 dex when compared to local low-metallicity galaxies, in contrast to the large offset in N/O of $-$0.32 dex that was found when KLEVER galaxies are compared to the full SDSS sample. This suggests that the production mechanisms for nitrogen in high-redshift galaxies are more similar to low-metallicity local galaxies.}
\label{fig:SDSS_KLEVER_NOOH_pri}
\end{figure}

To investigate the chemical enrichment history of the KLEVER galaxies further, we recall that the majority of these galaxies were found to lie below a metallicity of 12+log(O/H) $\sim$ 8.5, towards the region of the N/O-O/H plane where the N/O ratio plateaus at log(N/O) $\sim$ $-$1.5. In Figure \ref{fig:SDSS_KLEVER_NOOH_pri}, we plot the stellar mass-N/O relation for only those SDSS galaxies with 12+log(O/H) < 8.5. The contours represent the distribution of the SDSS galaxies, for which we take the average N/O in 0.25 dex intervals of stellar mass and fit a straight line, shown in black. The best fit for the total SDSS sample is shown as a red line. The magenta points show the KLEVER galaxies, and the histogram shows the deviation of SDSS and KLEVER galaxies from the best-fit black line. A linear regression is used for this best-fit as by limiting the sample to only those galaxies with low metallicities, we do not span a wide enough range of stellar masses to probe either the low or high mass plateaus in N/O, leaving only bins that probe the central linear region of the stellar mass-N/O plane. It is clear from this plot that KLEVER galaxies behave similarly to low-metallicity local galaxies in this plane, which suggests that the same mechanisms govern the production of nitrogen in high-$z$ galaxies and low-metallicity local galaxies. Within the SDSS galaxies, a clear dependence on stellar mass is observed, which suggests that even within the low-metallicity galaxies which lie towards the low N/O plateau variable contributions from secondary N production mechanisms, or delayed N enrichment, may be present. 

Even without considering the star formation rate, this conclusion is in agreement with our interpretation of the fundamental nitrogen relation. If KLEVER sources are young galaxies with high star formation efficiencies, then they will be able to attain a large stellar mass and metallicity in a short time, with an enrichment history dominated by the products of core-collapse supernovae, in agreement with the results of other works studying high-$z$ galaxies (\citealt{Steidel_2016}; \citealt{Strom_2018};\citealt{Shapley_2019}; \citealt{Sanders_2020OH};  \citealt{Topping_2020},\citeyear{Topping_2020b}; \citealt{Cullen_2021}).The delayed enrichment of nitrogen relative to oxygen, mainly caused by low and intermediate mass stars, can then cause the N/O ratio to increase, with N/O increasing to greater values in the more massive, metal-rich galaxies due to a greater impact from secondary N production in low and intermediate mass stars. To confirm the feasibility of the scenario described above, we plot a galaxy evolutionary track from the chemical evolution models of \cite{Vincenzo_2016} (see also \citealt{Magrini_2018}) in Figure \ref{fig:SDSS_KLEVER_NOOH_pri}. We note that the model used assumes a \cite{Salpeter_1995} IMF, although we expect that a change to a \cite{Chabrier_2003} IMF would have little effect on the shape of the evolutionary track, with a mild change in the normalisation with stellar mass. The evolutionary track is characterised by a large star formation efficiency of 5 Gyr$^{-1}$, with the track moving horizontally on the plane, increasing stellar mass without enriching N/O, before N/O rapidly begins to increase towards the average relation for the full sample. In Figure~\ref{fig:SDSS_KLEVER_NOOH_pri} we highlight where the galaxy would lie on the plane at an age of 1 Gyr. A galaxy on the evolutionary track with log(N/O) $\sim$ -1.2, roughly the median value of the KLEVER sample, would have an age of $\sim$ 0.5 Gyr. This suggests that some of the scatter in N/O values could be attributed to differences in galaxy age (or star formation history), with galaxies with mass-weighted ages that are older having had more time for nitrogen enrichment to occur. More detailed modelling of nitrogen abundances is the topic for future work, however if variation in star formation history of galaxies can explain the scatter of N/O with stellar mass, then it may also be responsible for the large observed scatter in the N/O-O/H plane.

\section{Conclusions}\label{Conc}

In this paper, we have analysed the scaling relations between the nitrogen-over-oxygen (N/O) abundance and various galactic properties including metallicity, stellar mass and star formation rate. We introduced for the first time the full KLEVER sample of high redshift galaxies, and compared a sub-sample of these galaxies to the nitrogen abundance scaling relations found locally. We also introduced new direct-temperature based calibrations of N/O abundance using the N2O2 and N2S2 diagnostics, calibrated using the SDSS stacks presented in \cite{Curti_2017}. The main findings of the paper are summarised below. 

\begin{itemize}
  \item KLEVER galaxies have N2O2 values in good agreement with those observed by other authors at this redshift (e.g \citealt{Shapley_2015}; \citealt{Steidel_2016}; \citealt{Strom_2017}). When converted into an estimate of log(N/O) using the new calibrations presented in this paper, we see a marginal enhancement of 0.1 dex in nitrogen abundance, at a fixed metallicity, for our KLEVER sample relative to local galaxies. Given the large scatter of individual objects in the N/O-O/H plane, particularly towards lower metallicities, this mild offset suggests their is no strong evolution in the N/O-O/H relationship with redshift.
  \item At a fixed stellar mass, KLEVER galaxies have lower nitrogen abundances than local galaxies by an average 0.35 dex. This finding indicates a strong evolution in the stellar mass-N/O relation, in agreement with \cite{Strom_2017}. This evolution is primarily driven by the evolution of the mass-metallicity relation, with KLEVER galaxies having lower metallicities than SDSS galaxies when compared at a fixed stellar mass, and hence lower N/O values.
  \item KLEVER galaxies fall slightly below the Fundamental Metallicity Relation (FMR), as parameterised by \cite{Curti_2020FMR}, by an average of $\sim$ 0.1 dex in O/H when using local metallicity calibrations. This offset is within the uncertainties of using the FMR to predict O/H values, particularly when considering that the KLEVER galaxies lie in a region of the FMR extrapolated from the local FMR, suggesting that the equilibrium between gas-phase processes that govern the FMR locally is still in place at $z$ $\sim$ 2.
  \item Similar to the secondary dependence of the mass-metallicity relation on star formation rate, we find a secondary dependence on SFR within the stellar mass-N/O plane. This finding is different to the results found in \cite{Perez-Montero_2013} due to our adoption of N2O2 as a diagnostic for N/O rather than N2S2 (which we have shown to be affected by a dependence on SFR). The finding that N/O has a secondary dependence on SFR similar to O/H, suggests that the FMR is not purely driven by processes that only O/H is sensitive to, such as pristine gas inflows, but also processes that are probed by N/O, such as galaxy age or star formation history, that trace the chemical enrichment histories of galaxies. 
  \item Analogous to the FMR, we use the dependence on SFR within the stellar mass-N/O plane to construct a Fundamental Nitrogen Relation (FNR). We find that the reduced N/O at a fixed stellar mass for high-$z$ galaxies can be accounted for by an increase in star formation rate. This suggests that high redshift galaxies may be young and highly efficient at converting their gas into stars. The apparent fundamentality of the FNR suggests that processes relating to the chemical enrichment history of galaxies evolve smoothly from $z$ $\sim$ 2 to $z$ $\sim$ 0. Furthermore, we find that the measured FNR is near-identical to a model FNR, constructed using a combination of the fundamental metallicity relation and a redshift-invariant N/O-O/H relation, suggesting the FNR is a by-product of these relations.
  \item The trend of N/O with stellar mass for the KLEVER galaxies most closely resembles that of local galaxies with low metallicities, for which variable contributions from secondary nitrogen production may contribute to the N/O ratio. The location of KLEVER galaxies on this plane is consistent with a scenario in which they are young with high star formation efficiencies, allowing them to attain high stellar masses in a short time before the enhancement due to the secondary production becomes dominant.
  
\end{itemize}

\section*{Data Availability}
The raw KLEVER data underlying this article is publicly available
through the ESO science archive facility (\url{http://archive.eso.org/cms.html}), whilst the SDSS emission line data is available at \url{http://classic.sdss.org/dr7/}. All other data contributing to this article will be shared on reasonable request to the corresponding author.

\section*{Acknowledgements}

We gratefully acknowledge the anonymous referee for their constructive comments that helped improve the paper.

Based on observations made with ESO Telescopes at the
La Silla Paranal Observatory under programme ID 197.A0717(A). CH-P, MC and RM acknowledge support by the Science and Technology Facilities Council (STFC) and from ERC Advanced Grant 695671 “QUENCH”. C.K. acknowledges funding from the UK Science and Technology Facility Council (STFC) through grant ST/R000905/1 and ST/V000632/1, and the visitor programme at Kavli Institute for Cosmology, Cambridge. YP acknowledges National Science Foundation of China (NSFC) Grant No. 12125301, 11773001 and 11991052.

This work utilizes gravitational lensing models produced by PIs Bradač, Natarajan \& Kneib (CATS), Merten \& Zitrin, Sharon, Williams, Keeton, Bernstein and Diego, and the GLAFIC group. This lens modeling was partially funded by the HST Frontier Fields program conducted by STScI. STScI is operated by the Association of Universities for Research in Astronomy, Inc. under NASA contract NAS 5-26555. The lens models were obtained from the Mikulski Archive for Space Telescopes (MAST). Funding for the SDSS and SDSS-II has been provided by the Alfred P. Sloan Foundation, the Participating Institutions, the National Science Foundation, the US Department of Energy, the National Aeronautics and Space Administration, the Japanese Monbukagakusho, the Max Planck Society and the Higher Education Funding Council for England.

%%%%%%%%%%%%%%%%%%%%%%%%%%%%%%%%%%%%%%%%%%%%%%%%%%

%%%%%%%%%%%%%%%%%%%% REFERENCES %%%%%%%%%%%%%%%%%%

% The best way to enter references is to use BibTeX:

\bibliographystyle{mnras}
\bibliography{nitrogen_paper} % if your bibtex file is called example.bib

% Alternatively you could enter them by hand, like this:
% This method is tedious and prone to error if you have lots of references
%\begin{thebibliography}{99}
%\bibitem[\protect\citeauthoryear{Author}{2012}]{Author2012}
%Author A.~N., 2013, Journal of Improbable Astronomy, 1, 1
%\bibitem[\protect\citeauthoryear{Others}{2013}]{Others2013}
%Others S., 2012, Journal of Interesting Stuff, 17, 198
%\end{thebibliography}

%%%%%%%%%%%%%%%%%%%%%%%%%%%%%%%%%%%%%%%%%%%%%%%%%%

%%%%%%%%%%%%%%%%% APPENDICES %%%%%%%%%%%%%%%%%%%%%

\appendix

\section{KLEVER galaxy spectra}
\label{sec:KLEVER_spec}

\begin{figure*}

    \begin{subfigure}{0.48\textwidth}
      \centering
      % include first image
      \includegraphics[width=\linewidth]{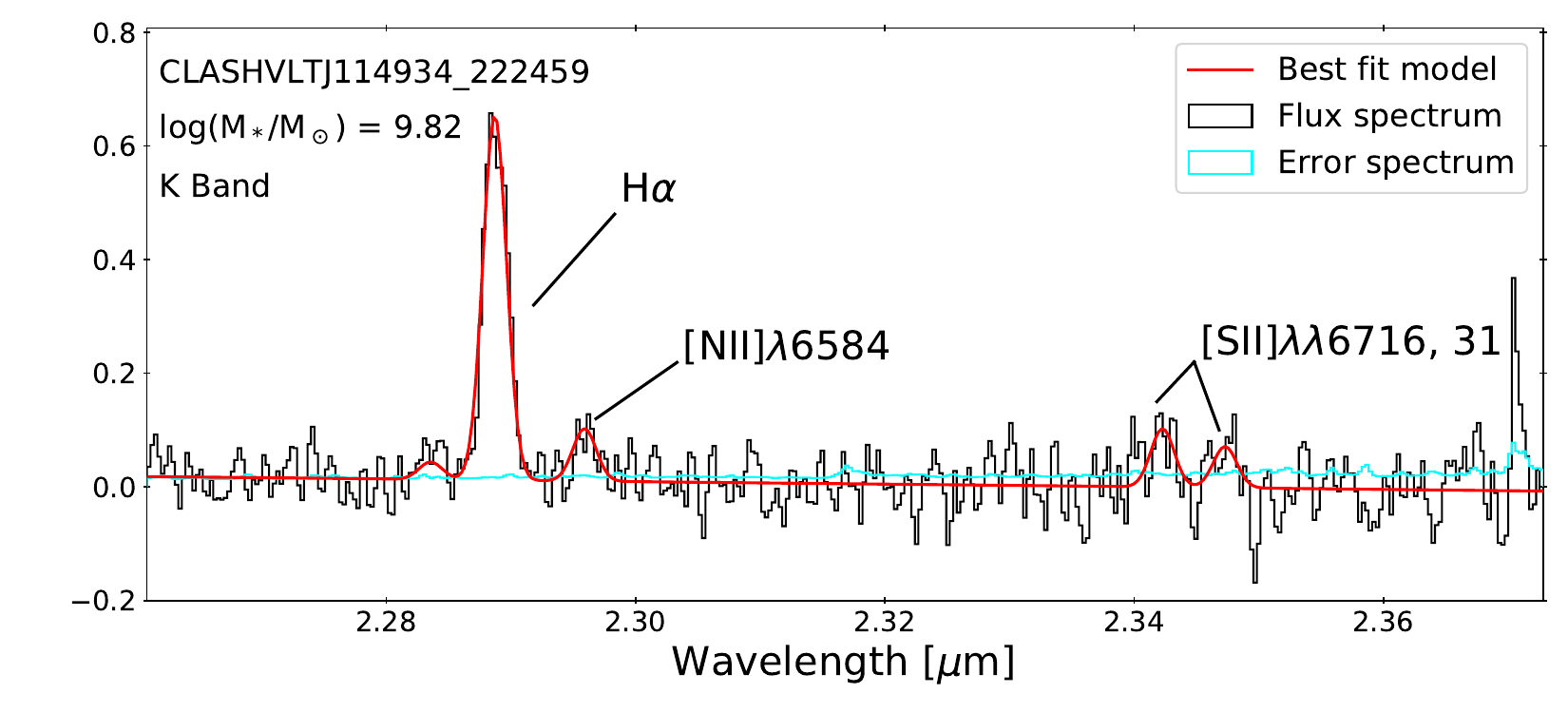}\par
      \vspace*{-4.2mm}
      \includegraphics[width=\textwidth]{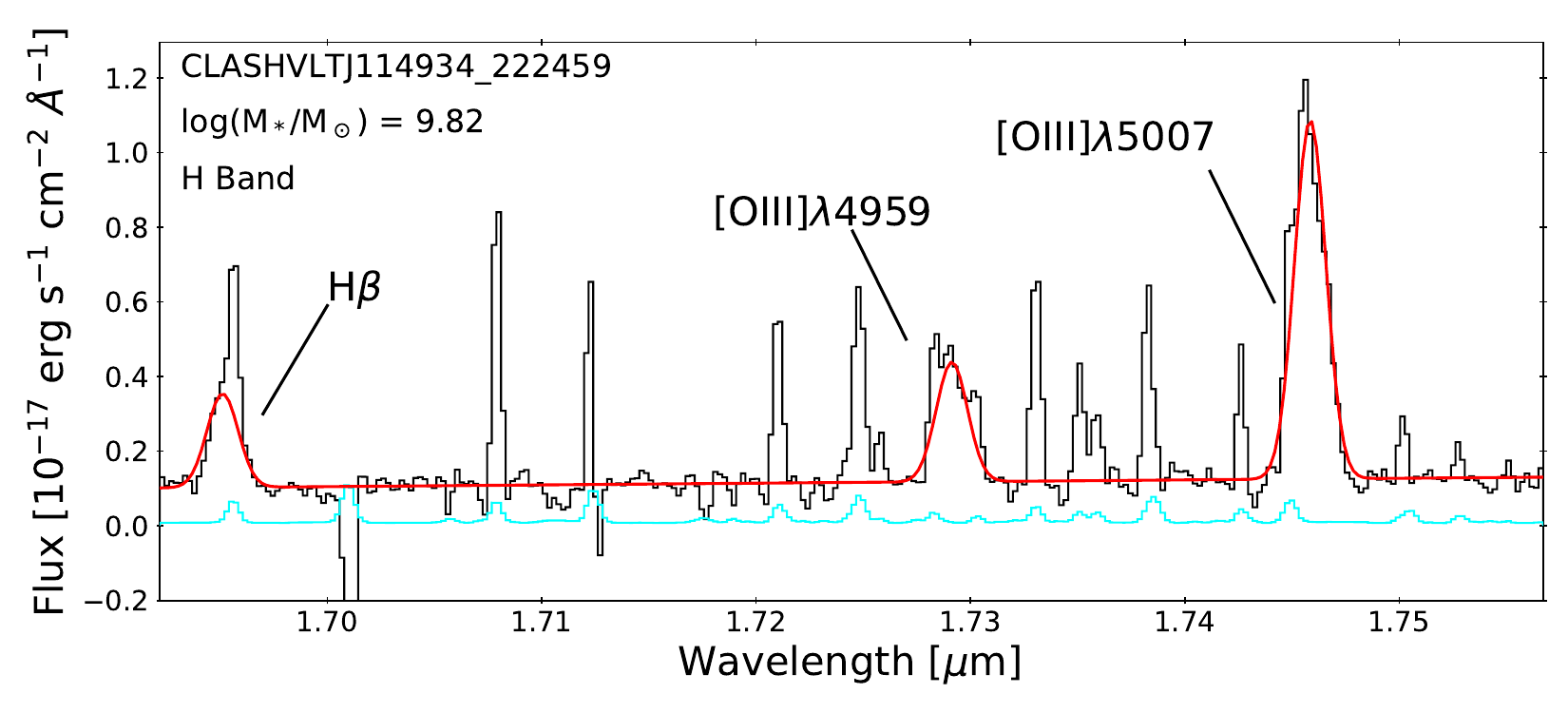}\par
      \vspace*{-4.2mm}
      \includegraphics[width=\textwidth]{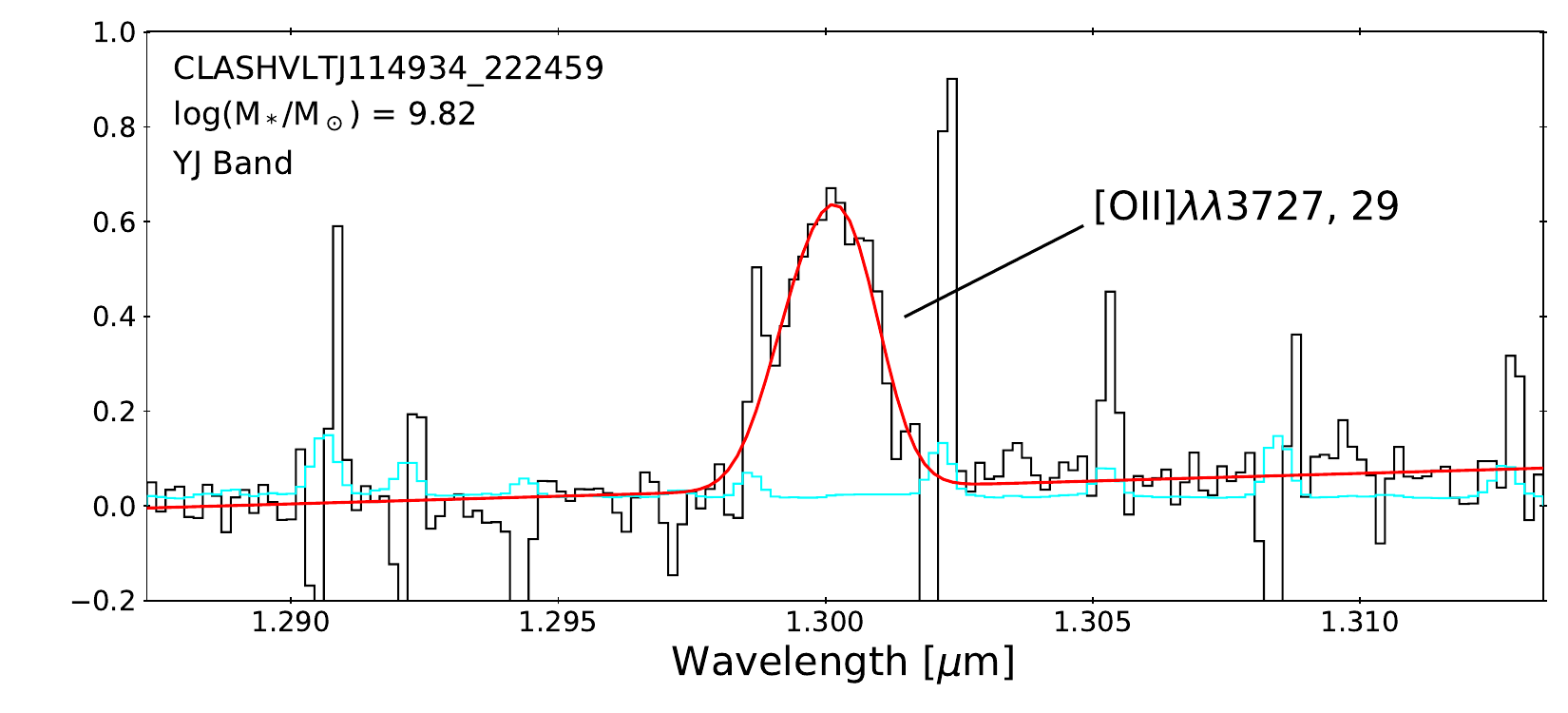}\par
      \centering
      \caption{CLASHVLTJ114934\_222459, lensed galaxy at $z\sim$2.49.}
      \label{fig:sub-macs}
    \end{subfigure}\quad
    \begin{subfigure}{0.48\textwidth}
      \centering
      % include first image
      \includegraphics[width=\linewidth]{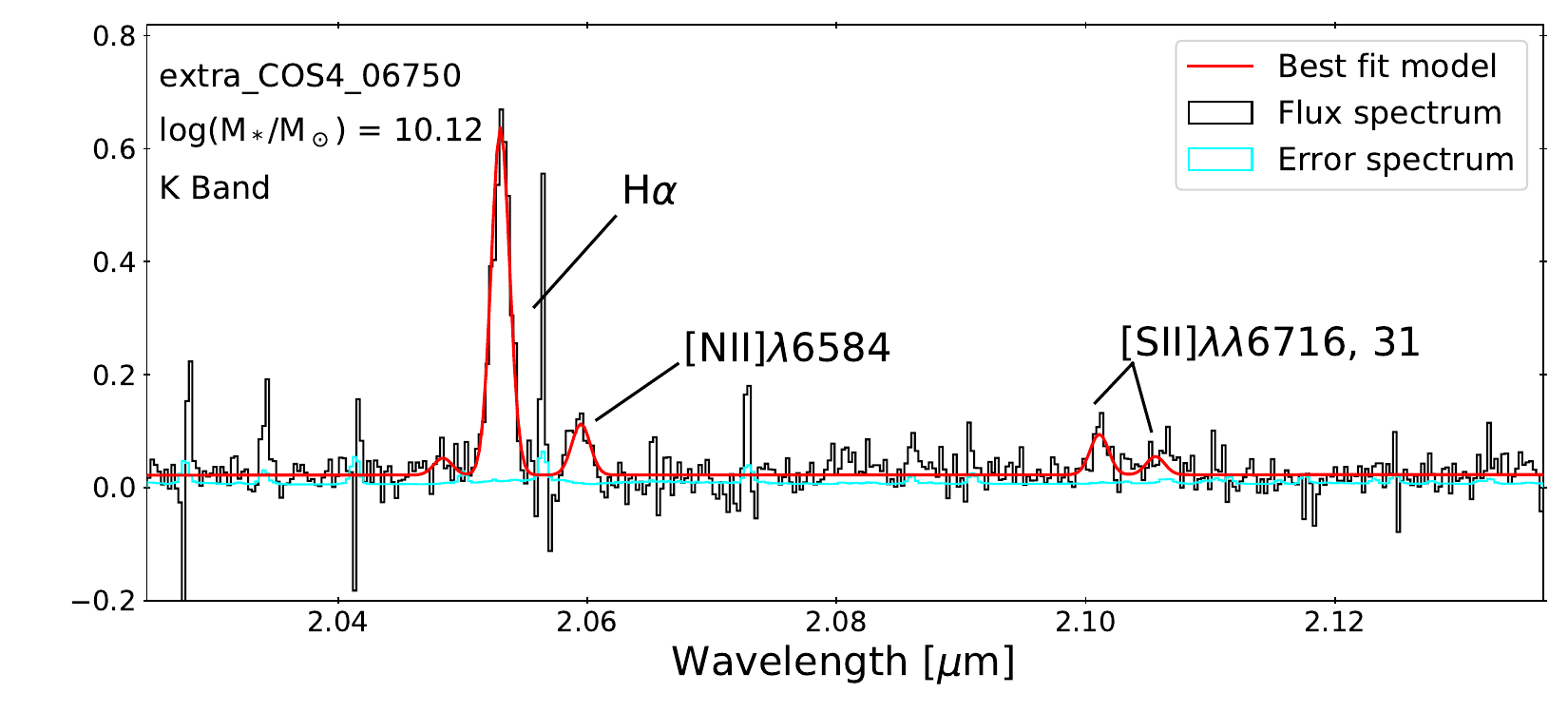}\par
      \vspace*{-4.2mm}
      \includegraphics[width=\textwidth]{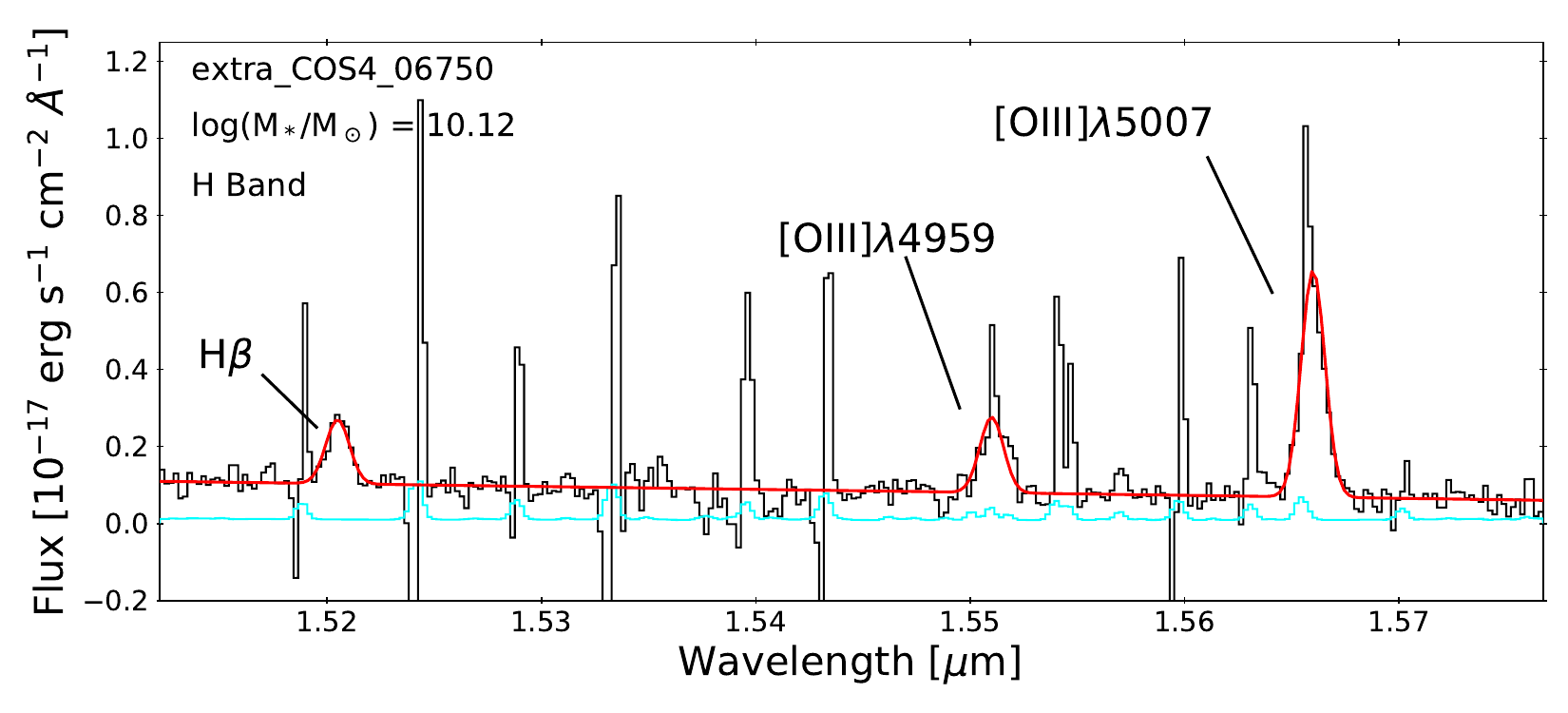}\par
      \vspace*{-4.2mm}
      \includegraphics[width=\textwidth]{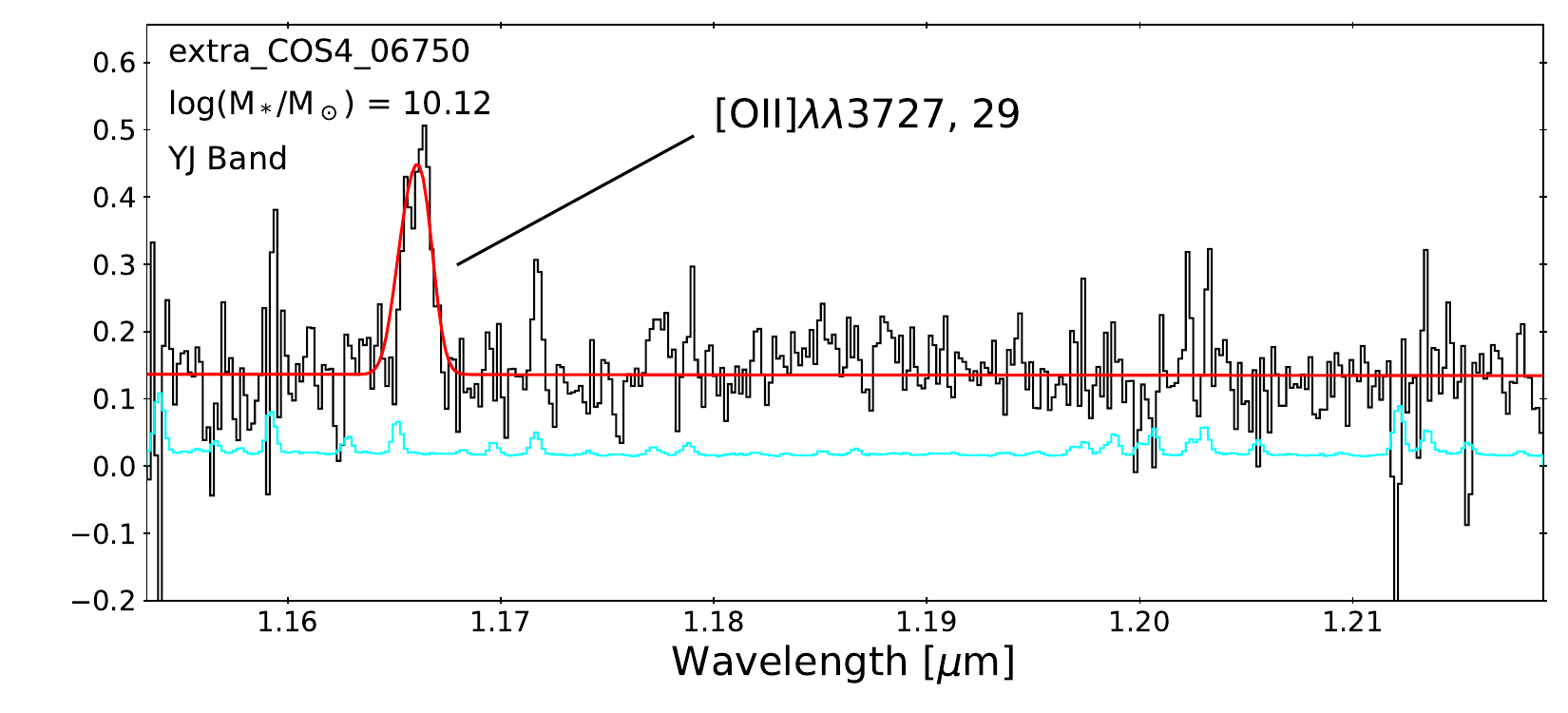}\par
      \caption{extra\_COS4\_06750, galaxy at $z\sim$2.13.}
      \label{fig:sub-cosmos}
    \end{subfigure}\quad
    \begin{subfigure}{0.48\textwidth}
      \centering
      % include first image
      \includegraphics[width=\linewidth]{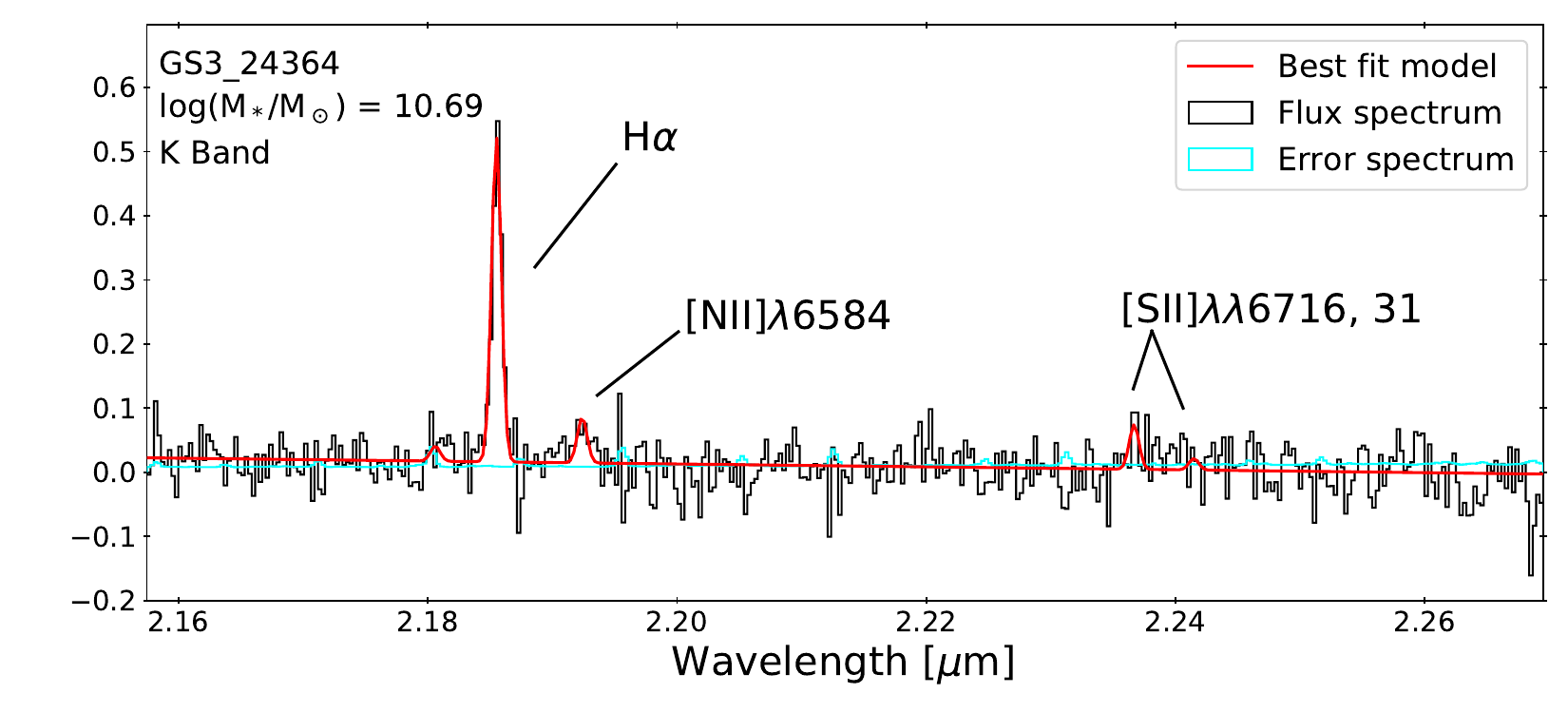}\par
      \vspace*{-4.2mm}
      \includegraphics[width=\textwidth]{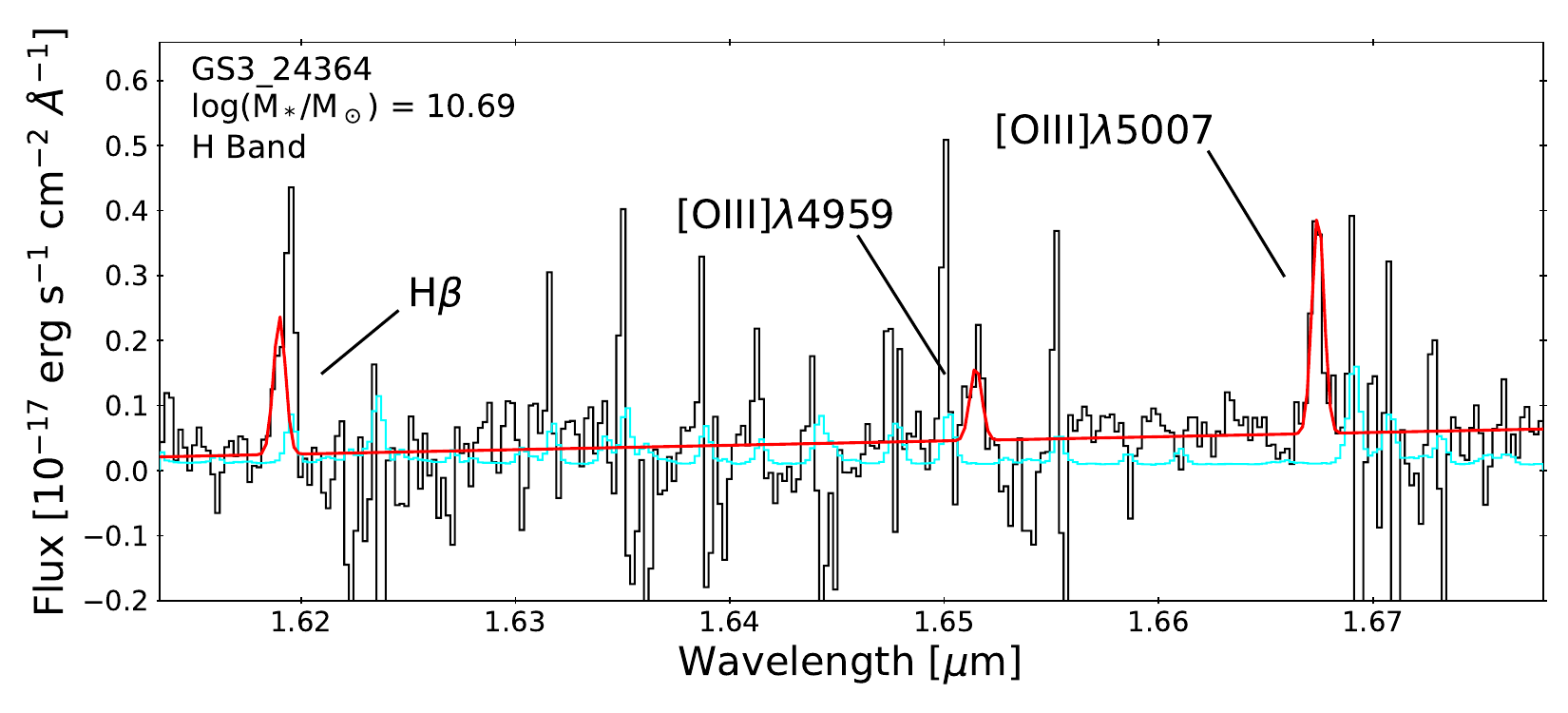}\par
      \vspace*{-4.2mm}
      \includegraphics[width=\textwidth]{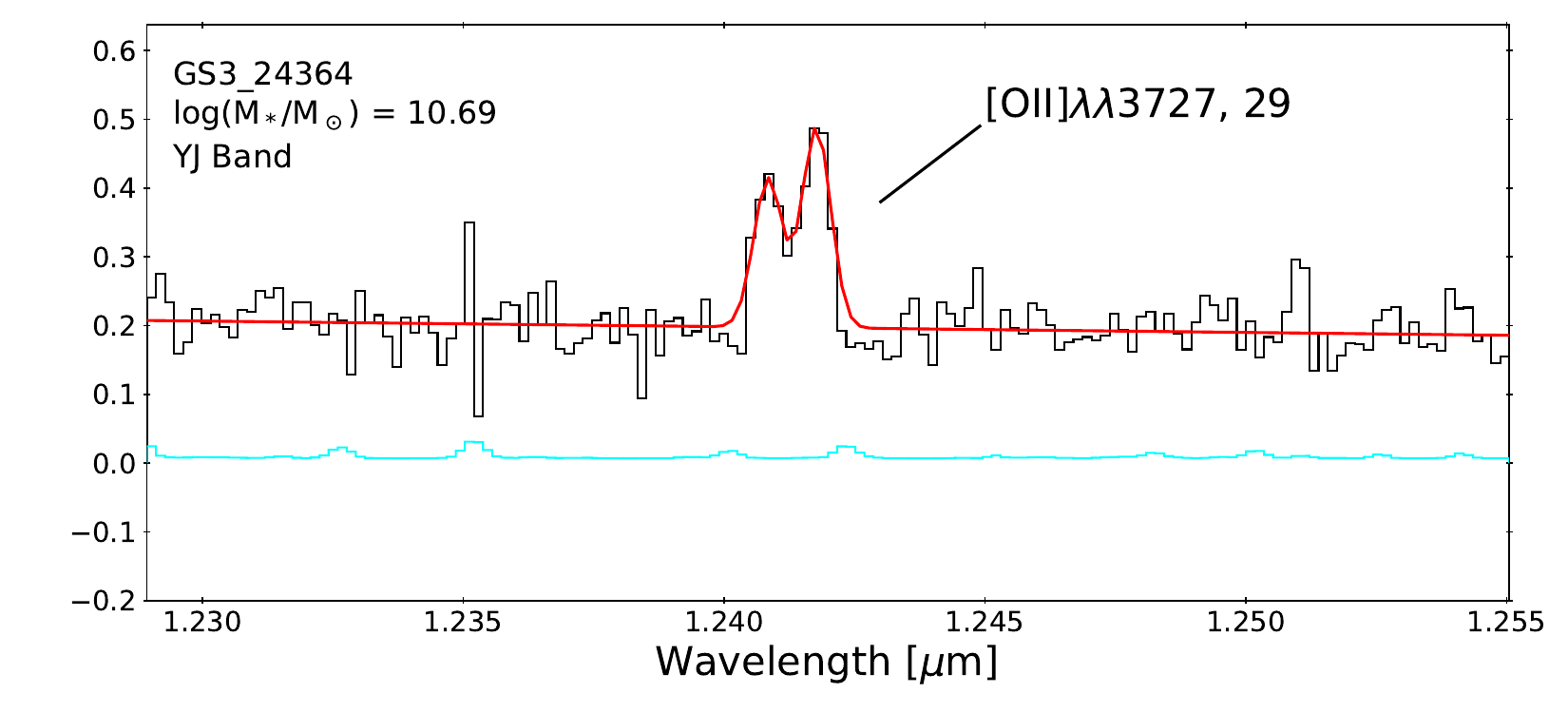}\par
      \caption{GS3\_24364, galaxy at $z\sim$2.33.}
      \label{fig:sub-goods}
    \end{subfigure}\quad
    \begin{subfigure}{0.48\textwidth}
      \centering
      % include first image
      \includegraphics[width=\linewidth]{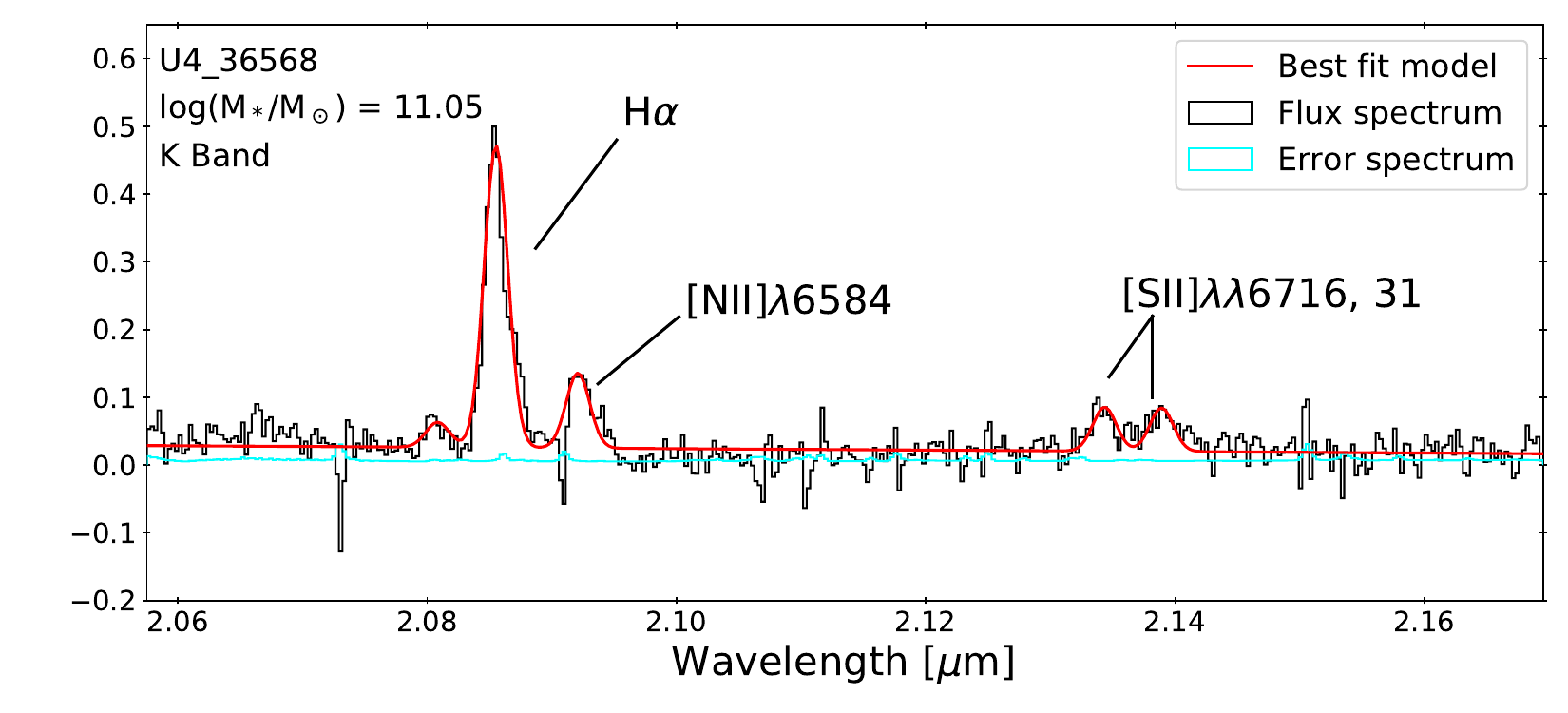}\par
      \vspace*{-4.2mm}
      \includegraphics[width=\textwidth]{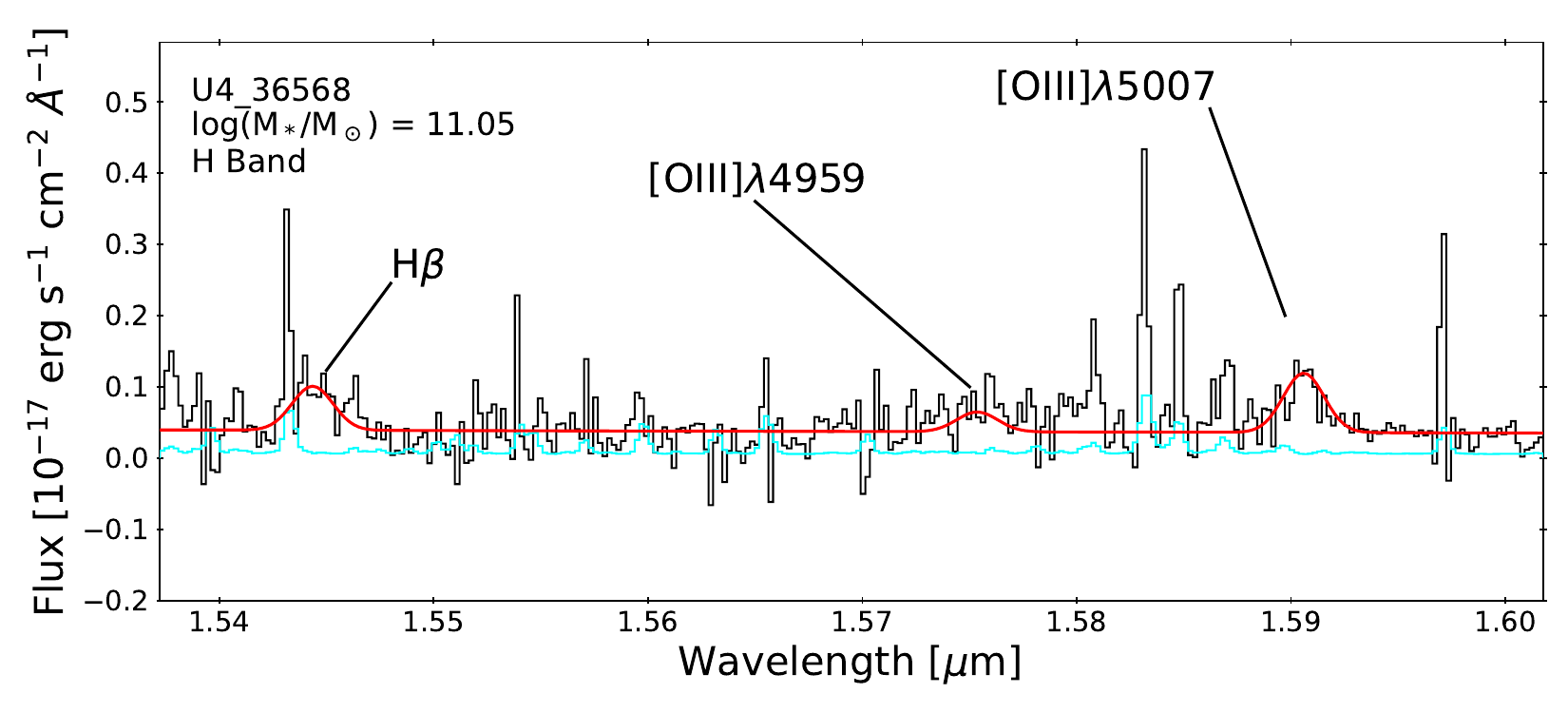}\par
      \vspace*{-4.2mm}
      \includegraphics[width=\textwidth]{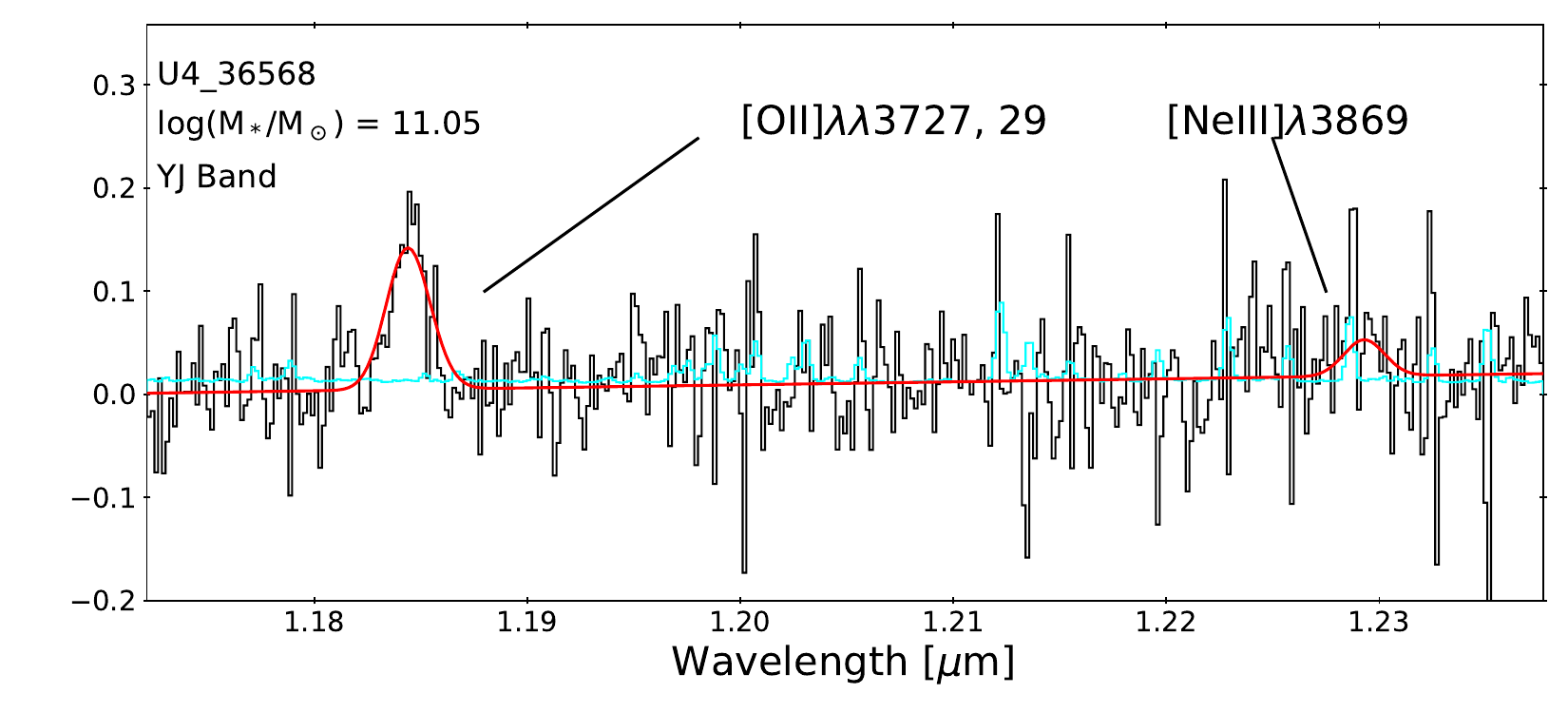}\par
      \caption{U4\_36568, galaxy at $z\sim$2.18.}
      \label{fig:sub-UDS}
    \end{subfigure}\quad

    \caption{Integrated spectra for 4 KLEVER galaxies in the K, H and YJ bands. The best-fit to the emission lines are shown as red lines, with the cyan lines showing the integrated error spectra.}
    \label{spectra}
\end{figure*}

\begin{figure*}

    \begin{subfigure}{\textwidth}
      \centering
      % include first image
      \includegraphics[width=\textwidth]{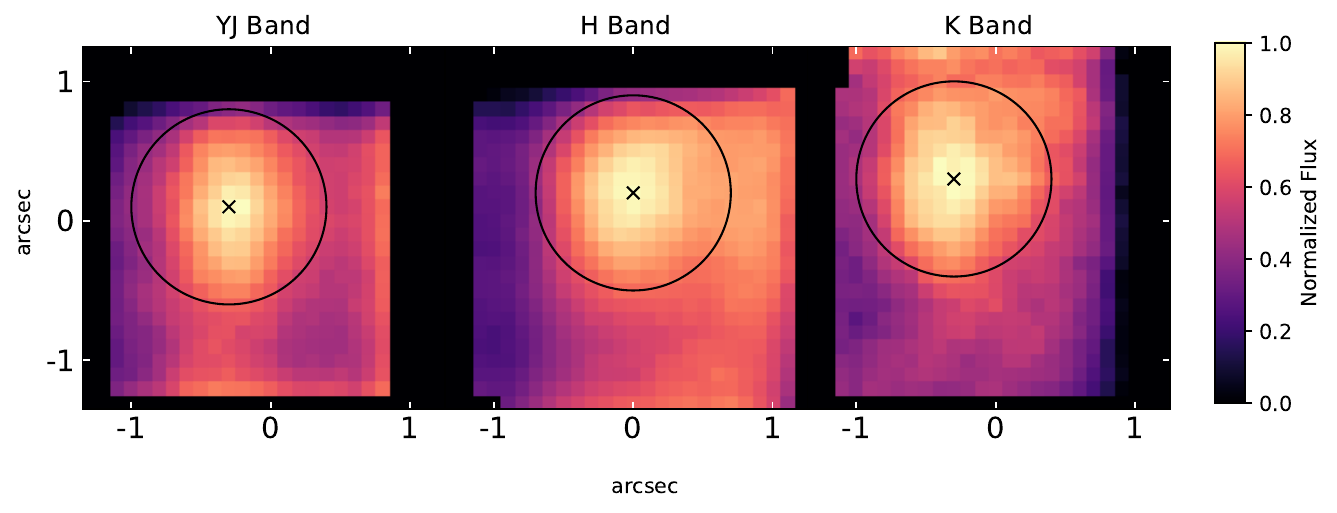}
      \centering
      \vspace*{-4.2mm}
      \caption{Normalized continuum fluxes in all 3 bands for the lensed object CLASHVLTJ114934\_222459.}
      \label{fig:sub-macs-app}
    \end{subfigure}
    
    \begin{subfigure}{\textwidth}
      \centering
      % include first image
      \includegraphics[width=\textwidth]{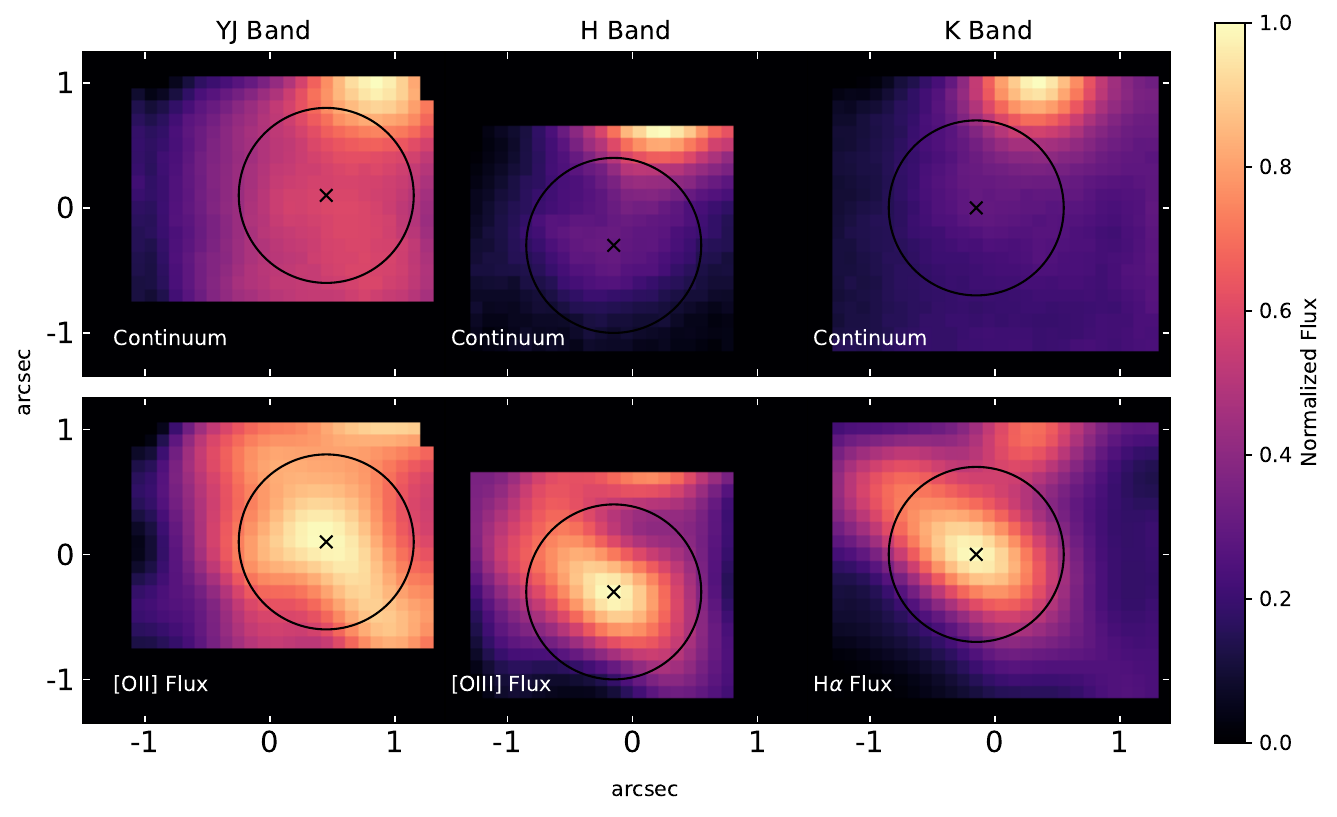}
      \centering
      \caption{Normalized continuum and emission line fluxes in all 3 bands for GS3\_24364.}
      \label{fig:sub-goods-app}
    \end{subfigure}

    \caption{Examples of the aperture placement used to extract the integrated flux spectra for the KLEVER sample. The origin of the coordinate system in each panel is the centre of the original datacube, which the aperture may be offset from if the observed galaxy is not located in the middle of the datacube. The black cross shows where the aperture was placed to extract the integrated spectra, whilst the black circle shows the region encompassed by the aperture.(a) Galaxy for which the continuum flux is detected in each band when collapsing over the full wavelength range whilst masking both emission lines and strong OH lines. (b) Galaxy for which the continuum is not clearly detected due to contamination by a nearby source. In this case, the aperture was centred on the peak of the strongest emission line in each band, shown in the bottom row. 
    }
    \label{aperture}
\end{figure*}

In Fig~\ref{spectra} we present the integrated emission spectra of 4 KLEVER galaxies taken from the sample analysed within this paper in the K, H and YJ bands. The best-fit models to this spectra are shown in red, with the integrated noise spectra shown in cyan. We present galaxies across a range of stellar masses (log(M$_*$/M$_{\odot}$ = 9.82, 10.12, 10.69 and 11.05 respectively). Three galaxies are un-lensed, taken from the COSMOS, GOODS-S and UDS fields, with the fourth galaxy being lensed by the MACS J1149.5+2223 cluster. Peaks in the integrated noise spectra are seen to correspond with regions that are contaminated by strong OH sky lines. It can be seen that even when there are OH sky lines very close to nebular emission lines, the weighting of the best-fit procedure by the noise spectrum means contamination by the sky lines is well avoided. The signal-to-noise for all of the major nebular emission lines probed by KLEVER measured for the four galaxies presented in Figure~\ref{spectra} are given in Table~\ref{tab:KLEVER_SN}. Examples of the aperture placement used to extract the integrated spectra are shown in Figure~\ref{aperture} for two different cases: firstly, the case where the continuum of the galaxy can be detected, and secondly the case where no continuum flux is detected, and instead the flux of the strongest emission line in each band is used to centre the aperture.

\begin{table}
 \centering
 \begin{tabular}{ccccc}
  \hline
  \hline
  Emission Line & (a) & (b) & (c) & (d)\\
  \hline
  [\ion{O}{II}]$\lambda\lambda$3727,29 S/N  &  14.7  &  17.6 & 18.3 & 9.7\\
  H$\beta$ S/N & 7.1 & 12.0 & 5.8 & 5.8\\\relax
  [\ion{O}{III}]$\lambda$5007 S/N & 41.2 & 34.1 & 9.5 & 8.4\\
  H$\alpha$ S/N &  32.1 & 44.7 & 21.7 & 43.4\\\relax
  [\ion{N}{II}]$\lambda$6584 S/N & 5.8 & 6.6 & 3.0 & 10.8\\\relax
  [\ion{S}{II}]$\lambda\lambda$6716,31 S/N & 2.4, 2.2 & 5.4, 2.4 & 2.7, 0.7 & 8.5, 8.4\\
  \hline
  \hline
 \end{tabular}
  \caption{The signal-to-noise of all of the major emission lines fit in the integrated spectra of our KLEVER sample for the four galaxies presented in Figure~\ref{spectra}, where (a) = CLASHVLTJ114934\_222459, (b) = extra\_COS4\_06750, (c) = GS3\_24364 and (d) = U4\_36568 . Note that the S/N presented for the [\ion{O}{II}]$\lambda\lambda$3727,29 doublet represents the flux measured for the total doublet as the individual emission lines are often blended due to their small wavelength separation, whereas for the [\ion{S}{II}]$\lambda\lambda$6716,31 emission doublet we report the S/N on the individual lines within the doublet.}
  \label{tab:KLEVER_SN}
\end{table}

\section{N/O calibrations with sSFR}
\label{sec:sSFR_appendix}

\begin{figure}
\includegraphics[width=\columnwidth]{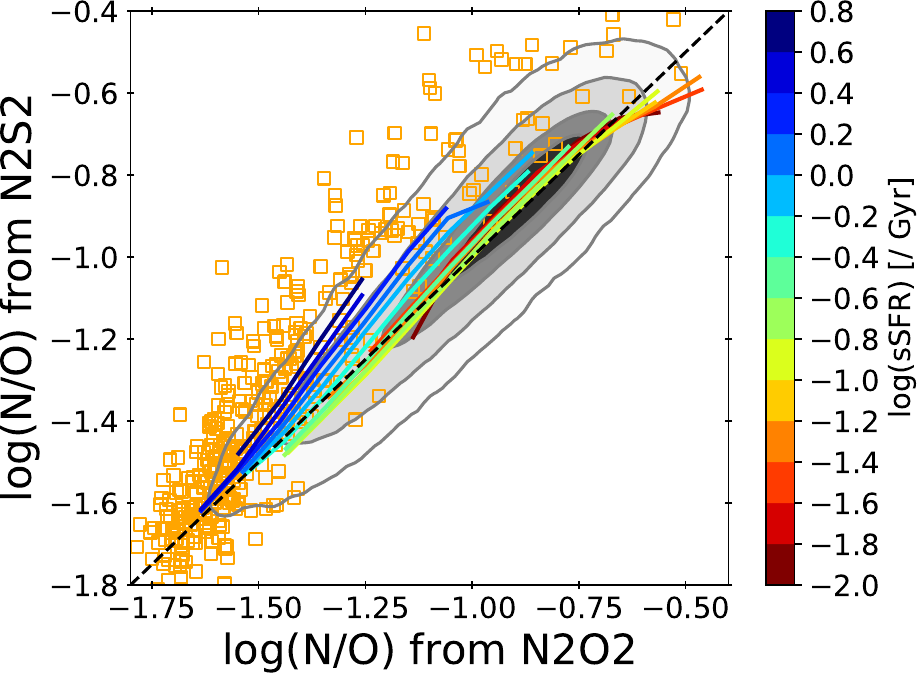}
\caption{N/O derived from both N2O2 and N2S2 for galaxies in our SDSS sample, with contours showing the regions encompassing 30\%, 60\% and 90\% of the total sample. The coloured lines show how the relationship between N2S2 and N2O2 varies as a function of sSFR. The orange squares show the location of local \textsc{Hii} regions, taken from \citet{Pilyugin_2012}.}
\label{fig:NO_diag_sSFR}
\end{figure}

To investigate further the differences that arise when using either N2O2 or N2S2 to estimate the N/O ratio, we estimate the nitrogen abundance from both of these diagnostic lines, using Equations \ref{N2O2_eq} and \ref{N2S2_eq} respectively, for our full SDSS sample. The results are shown as grey contours in Figure \ref{fig:NO_diag_sSFR}, overlaid with tracks of constant specific SFR (sSFR) taken in 0.2 dex wide sSFR bins. As can be seen, the low sSFR tracks all lie on the 1:1 relationship between N2O2 and N2S2, whilst the higher sSFR tracks begin to show a deviation in the N/O estimated from N2O2 compared to N2S2. This may suggest that the differences that arise when using N2O2 and N2S2 are driven by differences in the star formation activity within a galaxy. Galaxies with higher sSFR values that are more actively star forming may be dominated by large \textsc{Hii} regions, the central regions of which have greater ionisation parameters that allow preferential ionisation of higher ionisation species, such as S$^{++}$, compared lower ionisation potentials, such as S$^+$. The similarities in ionisation structure of O$^+$ and N$^+$ mean that observations of star forming galaxies dominated by these \textsc{Hii} regions would have elevated N2S2 values when considered at a fixed N2O2. For reference, local \textsc{Hii} regions taken from \cite{Pilyugin_2012} are shown as orange squares in Figure \ref{fig:NO_diag_sSFR}. As in the results of \cite{Strom_2018}, these \textsc{Hii} regions are offset from the 1:1 relation, which could be driven by a similarity in the ionisation conditions of these regions compared to high sSFR local galaxies. As shown in \cite{Mannucci_2021}, the clear offset between local galaxies and \textsc{Hii} regions may be a consequence of aperture effects, and if larger apertures were used to extract integrated spectra for the \textsc{Hii} regions the additional [\ion{S}{II}] measured in the outer emission regions would act to reduce the N2S2 values, producing a better agreement with local values.

%%%%%%%%%%%%%%%%%%%%%%%%%%%%%%%%%%%%%%%%%%%%%%%%%%

% Don't change these lines
\bsp	% typesetting comment
\label{lastpage}
\end{document}